\documentclass[journal,twocolumn,letterpaper]{IEEEJERM}
\ifCLASSINFOpdf
 \else
\fi

\usepackage{times,amsmath,epsfig}
\usepackage{fancyhdr}
\usepackage{amsmath}
\usepackage{amsfonts}
\usepackage{amssymb}
\usepackage[latin1]{inputenc}
\usepackage{array}
\usepackage{graphicx}
\usepackage{url}
\usepackage{subfigure}
\usepackage{overpic}
\usepackage{bm}
\usepackage{breqn}
\usepackage{xcolor}
\usepackage{soul}
\usepackage{amssymb}
\usepackage{xcolor}

\renewcommand{\v}[1]{\ensuremath{\mathbf{#1}}} 
\newcommand{\uv}[1]{\ensuremath{\mathbf{\hat{#1}}}} 
\newcommand{\pd}[2]{\frac{\partial #1}{\partial #2}} 
\newcommand{\pdd}[2]{\frac{\partial^2 #1}{\partial #2^2}} 

\begin{document}
\title{Frequency comb generation at 800nm in waveguide array quantum well diode lasers}
\author{Chang~Sun,
        Mark~Dong,~\IEEEmembership{Member,~IEEE,}
        Niall M. Mangan,
        Herbert G. Winful,~\IEEEmembership{Fellow,~IEEE,}
        Steven T. Cundiff,~\IEEEmembership{Fellow,~IEEE,}
        and J. Nathan Kutz
}

\twocolumn[
\begin{@twocolumnfalse}
  
\maketitle

\begin{abstract}
A traveling wave model for a semiconductor diode laser based on quantum wells is presented as well as a comprehensive theoretical model of the lasing dynamics produced by the intensity discrimination of the nonlinear mode-coupling in a waveguide array.  By leveraging a recently developed model for the detailed semiconductor gain dynamics, the temporal shaping effects of the nonlinear mode-coupling induced by the waveguide arrays can be characterized.  Specifically, the enhanced nonlinear pulse shaping provided by the waveguides are capable of generating stable frequency combs wavelength of 800 nm in a GaAs device, a parameter regime not feasible for stable combline generation using a single waveguide. Extensive numerical simulations showed that stable waveform generation could be achieved and optimized by an appropriate choice of the linear waveguide coupling coefficient, quantum well depth, and the input currents to the first and second waveguides. The model provides a first demonstration that a compact, efficient and robust on-chip comb source can be produced in GaAs.
\end{abstract}


\end{@twocolumnfalse}]

{
  \renewcommand{\thefootnote}{}%
  \footnotetext[1]{C. Sun, N. M. Mangan and J. N. Kutz are with the University of Washington, Seattle,
WA 98195 USA (e-mail: sunch610@uw.edu; niallmm@gmail.com; kutz@uw.edu).}
   \footnotetext[2]{M. Dong, S. T. Cundiff, and H. G. Winful are with the University of Michigan, Ann Arbor, MI 48109 USA (e-mail: markdong@umich.edu; 
cundiff@umich.edu; arrays@umich.edu).}
   \footnotetext[3]{M. Dong is with The MITRE Corporation, 202 Burlington Rd. Bedford, MA 01730 USA (e-mail: mdong@mitre.org). The author's affiliation with The MITRE Corporation is provided for identification purposes only and is not intended to convey or imply MITRE's concurrence with, or support for, the positions, opinions, or viewpoints expressed by the author.}
   \footnotetext[4]{Approved for Public Release; Distribution Unlimited. Public Release Case Number 19-0458.}
}
 
\IEEEpeerreviewmaketitle

\section{Introduction}

The ability to generate optical frequency combs on chip scale devices remains an open engineering challenge.   Such chip scale devices have the potential to revolutionize ultrafast and nonlinear optics for applications in frequency metrology and optical spectroscopy~\cite{Cundiff2003,Udem1999}, multi-heterodyne spectroscopy \cite{Coddington2008}, optical atomic clocks~\cite{Diddams2001}, and arbitrary waveform synthesis~\cite{Cundiff2010}.   A major obstacle in generating short pulses in diode lasers stems from the nonlinear phase shifts that occur due to fast carrier dynamics~\cite{Delfyett1992}, essentially limiting the pulse width inside the cavity. However, single-section diode lasers without saturable absorbers can also operate in a multimode phase-synchronized state known as frequency-modulated (FM) mode locking~\cite{Tiemeijer1989}.   A recent theoretical study developed a detailed traveling wave model of a semiconductor diode laser based on quantum wells~\cite{MarkIEEE}.   The resulting  frequency modulated comb showed potential for a compact, chip-scale comb source without additional external components at wavelength of 1550 nm.
However, the generation of a stable waveform was no longer producible at 
800 nm where many engineering applications are relevant. 
In this work, we modify the detailed physics model introduced by Dong et al.~\cite{MarkIEEE} to include coupled waveguide arrays (WGAs) in order to generate stable waveforms at 800nm.  WGAs have been demonstrated to enhance the nonlinear pulse shaping necessary to promote mode-locking~\cite{WilliamsWGA}, thus they are used in the present model to promote stable waveform generation at 
800nm. Indeed, we show that with the addition of the WGAs, stable waveform generation can be achieved in a diode laser configuration, allowing for the possibility of chip scale frequency combs.

Current methods for comb generation include the mode-locking of Ti:Sapphire laser \cite{Sutter1999} and fiber lasers \cite{Fermann2013}, as well as parametric frequency conversion due to the Kerr nonlinearity in passive microresonators \cite{Herr2012}. These approaches, however, require many discrete optical or fiber components, careful alignment, and bulky pump lasers and amplifiers, thus limiting their general utility outside of laboratories. 
Mode-locked diode lasers offer a portable and efficient solution for this technology by offering the direct generation of frequency combs from a chip-scale device \cite{Moskalenko2017, Rosales2011}. Typically, passively mode-locked diode lasers comprise two sections: a gain section and a reverse-biased saturable absorber section that leads to the formation of a periodic train of short pulses and hence a comb in the frequency domain. As noted, nonlinear phase shifts that occur due to fast carrier dynamics limit the minimum pulse width in the cavity \cite{Delfyett1992}.   However, single-section diode lasers without saturable absorbers can also operate in a multimode phase-synchronized state known as frequency-modulated (FM) mode locking \cite{Tiemeijer1989}. In the ideal FM mode locked state, the output is a continuous wave in time but the frequency modulation results in a set of comb lines with a fixed, non-zero phase difference. Such FM modelocked operation has been studied most intensively in quantum dot (QD) \cite{Gioannini2015, Rosales2012} and quantum dash \cite{Rosales2012-2} (QDash) lasers, but has also been observed in quantum well (QW) \cite{Sato2003, Calo2015} and bulk semiconductor lasers \cite{Tiemeijer1989}.

There have been many models published for semiconductor quantum well lasers with varying degrees of complexity. The simplest models include only a single rate equation and photon density variable \cite{Homar1996, Arakawa1986}, while more complex models may use multiple rate equations and more complex forms of the material polarization \cite{KN2010, McDonald1995, Jones1995, Vandermeer2005, Gordon2008, Lenstra2014} with varying degrees of phenomenological expressions and constants inserted. However, the existing models are usually insufficiently detailed to explain why FM combs arise in some QW lasers and not others, nor do they indicate which parameters need to be optimized for comb generation. The difficulty in modeling these types of diode lasers stems from the need to properly account for the many nonlinear effects in the semiconductor laser cavity.  Initial investigations on QD single-section lasers \cite{Gioannini2015} were recently more fully developed following previous works~\cite{Chow2002, Gioannini2015} in order to more fully characterize the FM comb generation in QW diode lasers~\cite{MarkIEEE}.  At 
1550 nm, this model was able to produce stable waveforms and excellent performance characteristics.  But in the application important 
800 nm parameter regime, the model was not capable of stable waveform generation.

The inability to produce FM combs in such a detailed QW diode laser model at 
800 nm motivates our exploration of 
using WGAs to enhance nonlinear pulse shaping for mode-locking and stabilize wave generation. Nonlinear mode-coupling in WGAs has been shown to produce the intensity discrimination necessary to produce mode-locking~\cite{Proctor05,Kutz08,Ching12}.  Indeed, semiconductor WGAs studied for CW lasing~\cite{winful1,winful2} motivated a recent phenomenological model for mode-locking on a chip~\cite{WilliamsWGA}.   Here, we integrate the detailed semiconductor physics model of Dong et al.~\cite{MarkIEEE} into a WGA in order to achieve stable waveform generation at 
800 nm. Indeed, we show that the extra pulse shaping induced by the WGAs are capable of producing frequency combs in this important parameter regime.

The paper is outlined as follows:  Sec.~\ref{sec:theory} details the theoretical model constructed for characterizing the electric field evolution dynamics and complex semiconductor gain physics.  Section~\ref{sec:results} shows detailed simulations of the full model and the resulting wave form generation.  The paper is concluded in Sec.~\ref{sec:conclusion} with a brief summary of our results.

\section{Theoretical models}
\label{sec:theory}

The following subsections outline the key modeling 
components necessary for constructing a detailed description of the laser cavity dynamics.  The interplay of the various physics allow for the generation of stable waveforms.

\subsection{Waveguide Arrays}
\begin{figure}[htbp]
\centerline{\includegraphics[width=\linewidth]{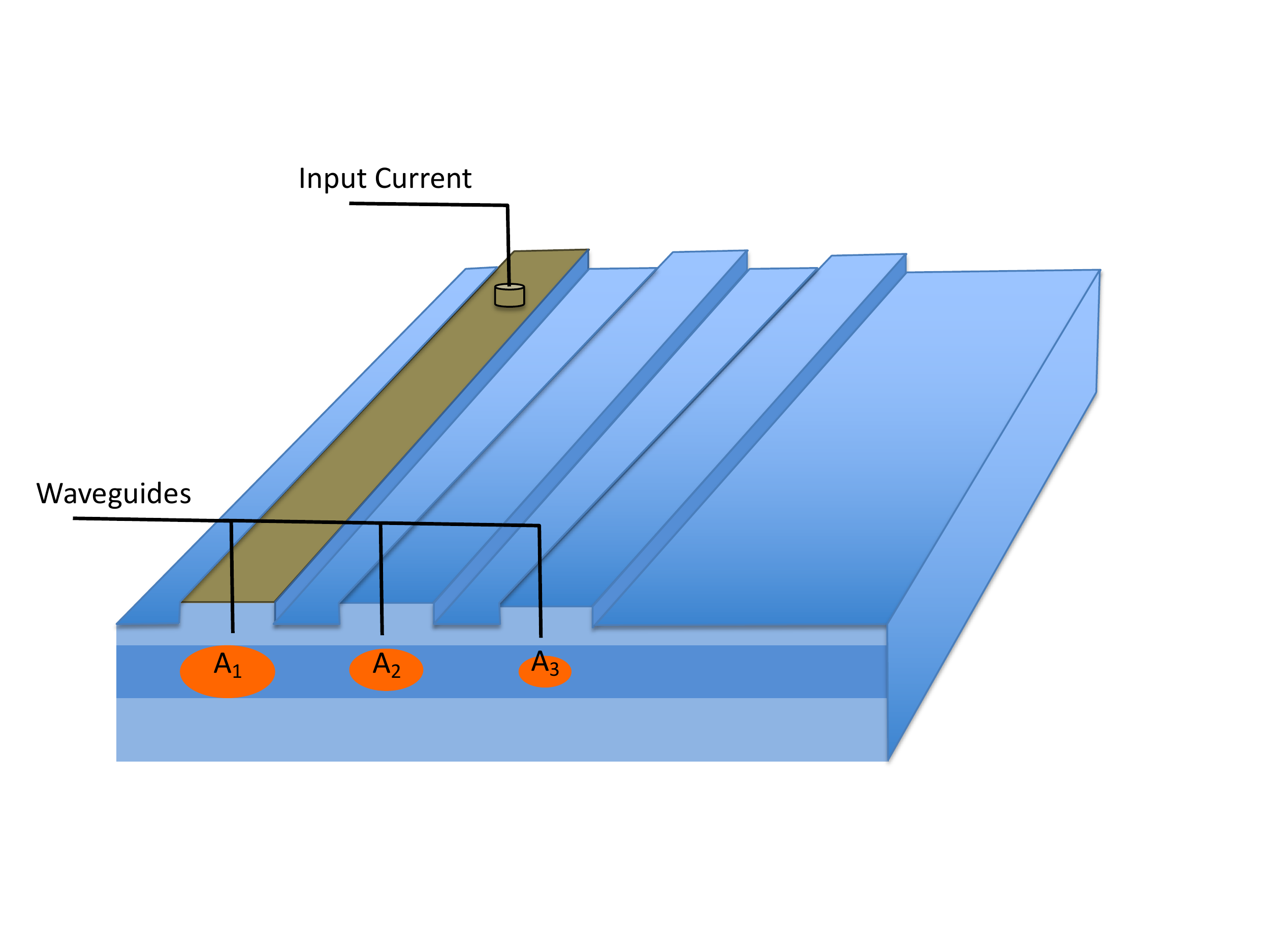}}
\caption{Schematic of a waveguide array diode laser. The input current on the first waveguide provides with the saturable absorption and amplification while the following waveguides two and three are for intensity discrimination and pulse shaping. Only waveguide one has net gain, the following waveguides two and three experience net losses.}
\label{schematic_WGA}
\end{figure}

The propagation of light in WGAs contributes to the self-focusing therefore enhances saturable absorber for mode-locking. 
This concept of a waveguide array mode-locked laser is shown in Figure \ref{schematic_WGA}. By preferentially coupling out 
low-intensity light to the neighboring waveguides, the electric field propagating in the first waveguide is shaped according to the intensity, dispersion and gain dynamics.   This intensity discrimination is necessary for the generation of stable mode-locked pulses in a laser cavity~\cite{Proctor05,Kutz08,Ching12}.  

The leading-order equations governing the electric field dynamics and the linear, evanescent electric field coupling to the neighboring waveguides are
given by 
\begin{equation}
i\frac{dA_{n}}{d\xi}+C(A_{n-1}+A_{n+1})+\beta|A_{n}|^{2}A_{n}=0,
\end{equation}
where $A_{n}$ represents the electric field envelope in the $nth$ waveguide in the array, $C$ represents the linear coupling coefficient, and $\beta$ the nonlinear self-phase modulation parameter \cite{christo,roberto,jstqe09_wga}. 

Though the waveguide array can be formed by a different numbers of waveguides, numerical studies and stability analysis shows that using three or more waveguides can produce robust pulse shaping and intensity discrimination~\cite{jstqe09_wga}. Indeed, a three waveguide structure has almost identical properties to the 41 waveguides considered in early WGA experiments~\cite{roberto}, whereas two waveguides do not offer stable and robust dynamics~\cite{jstqe09_wga}.  Therefore, we consider a WGA architecture with three waveguides.  The first waveguide is forward biased and gets a net gain from the pump.  
The second waveguide is weakly forward biased to be close to transparency such that it experiences a net linear loss. 
Waveguide three is reversely biased to increase the linear loss and shorten the recovery lifetime by sweeping out optically generated carriers, thus engineered to attenuate the electromagnetic energy that enters it.

A traveling wave model for the generation of stable frequency combs of a semiconductor diode laser in a single waveguide has been studied previously \cite{MarkIEEE}. Here we extend this model to GaAs, which has a higher central transition energy.  GaAs generates frequency combs around 
800 nm compared to 
1550 nm in \cite{MarkIEEE}. However, we show that the single waveguide model does not yield stable waveform generation in this new parameter regime.    The WGA architecture is a pulse shaping strategy that can help promote stable waveform generation.  This is achieved by using the intensity discrimination of the WGAs. 

\subsection{Gain model and governing equations}

\begin{figure}[htbp]
\centerline{\includegraphics[width=\linewidth]{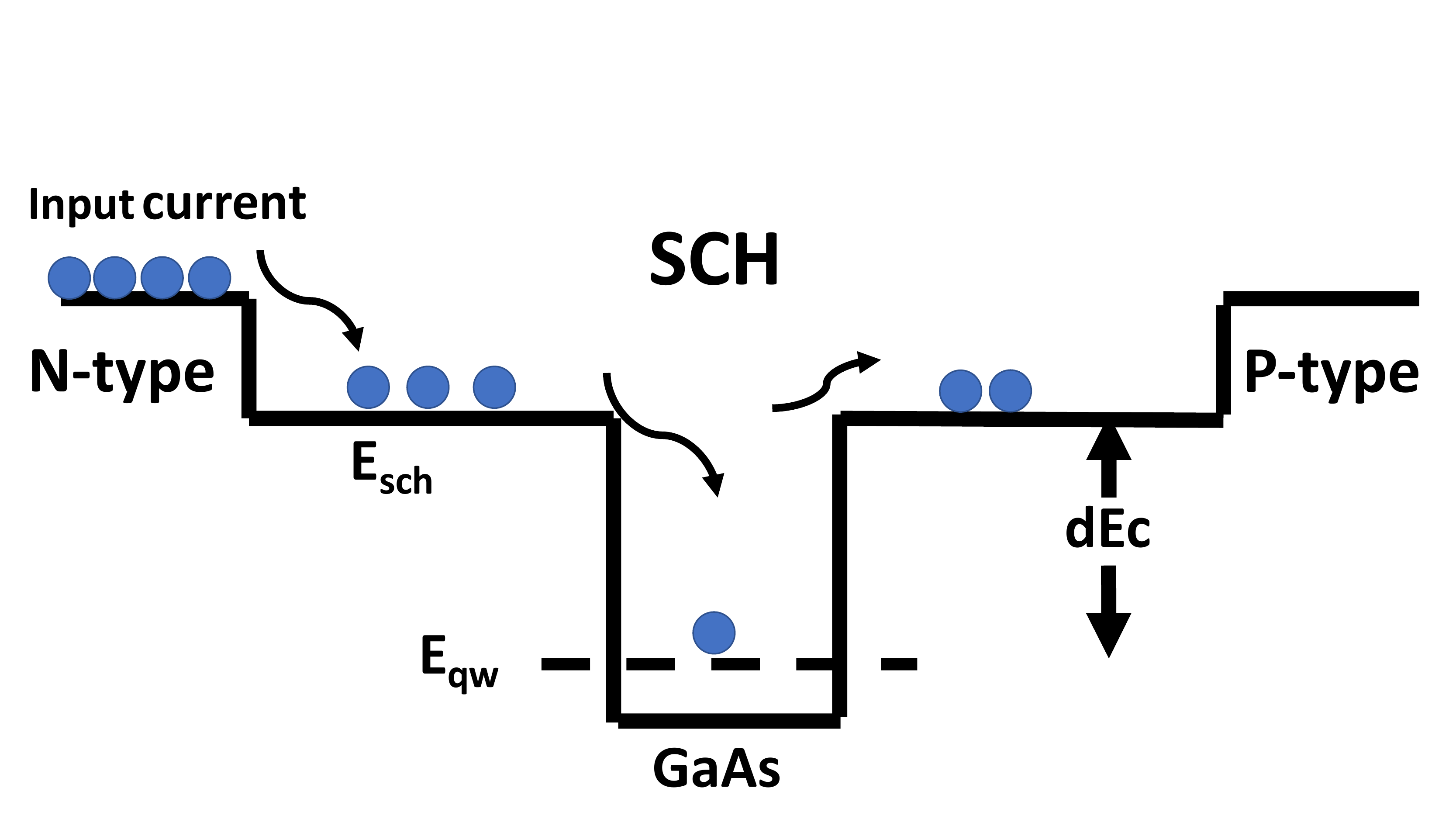}}
\caption{Schematic of a quantum well laser diode. Electrons can be trapped into the quantum well from the input current injection, and they have a 
probability of escaping from the quantum well as well.}
\label{wg}
\end{figure}

Here we describe the non-trivial gain model \cite{MarkIEEE}, and extend it to the three waveguide array.  A schematic is shown in Figure \ref{wg}. Electrons injected from the n side (holes from the p side) relax to the separate confinement heterostructure (SCH) layer, and become trapped in the quantum well. The most important difference between our quantum well model and previous models is that, for the carriers trapped in the quantum well, we have discretized the carrier equations in energy space and combined them with a multimode wave equation. While this approach increases the number of carrier equations to solve, it captures all the important dynamics of the multiple Fabry-Perot cavity modes and their interactions with carriers at different transverse energies. 

In a semiconductor, the carriers are typically confined in some type of nanostructure, such as a 2-D quantum well, a 1-D quantum wire or a 0-D quantum dot or dash, with an energy distribution determined by the $N$-dimensional density of states $D_r^{N-D}$ and occupation probability for electrons ($e$) or holes ($h$) $\rho^{e,h}$. We assume that the microscopic coherence decays sufficiently quickly such that each individual carrier emits light in a characteristic Lorentzian spectral lineshape with a homogenous linewidth $2\Gamma$ as determined by intraband relaxation effects. However, each group of carriers will emit at a different central frequency. In quantum wells in particular, the carriers have momenta in the unconfined directions that we quantify as the transverse energy $E_t$, and it is these energies that modify the transition frequency for all carriers with energy $E_t$. By integrating all carrier Lorentzians in energy space for each quantum well confined state, we have a gain term that accounts for homogenous and inhomogenous broadening, the asymmetric nature of the gain due to occupation levels and density of states, and the carrier-induced refractive index change. These complex Lorentzians also offer a simple way to calculate the real and imaginary parts of the gain without resorting to the Kramers-Kronig relations.

The electric field of the light wave in the cavity is taken as a sum of forward and backward components
\begin{align}
E(z,t) = E_+(z,t) e^{ik_0 z}+E_-(z,t)e^{-i k_0 z}
\end{align}
whose amplitudes satisfy the slowly-varying envelope equation
\begin{align}
\label{tw_eqn}
\pm\pd{}{z}E_\pm(z,t) + \frac{1}{v_g} \pd{}{t}E_\pm(z,t) = \Gamma_{xy} \frac{\omega_0^2}{2i k_0 c^2 \epsilon_0} \langle P_{tot}(t) e^{\mp ik_0z} \rangle
\end{align}
where the angular brackets signify averaging over a few wavelengths. Here, $v_g = c/n_0$ is the group velocity, $n_0$ is the group refractive index, $\Gamma_{xy}$ is the transverse confinement factor, $\omega_0$ is the central photon frequency (the choice of $\omega_0$ can be arbitrary but is generally chosen to be the transition frequency at the band edge), and $k_0 = n_0 \omega_0/c$. With Bloch Equations tailored to semiconductors as well as the standard adiabatic approximation (see Appendix), the total polarization for a 2-D quantum well is:
\begin{eqnarray*}
\label{ptot_sumk}
P_{tot}(t) &&= \frac{2}{V}  \sum_{\v{k}}  d^*_{ev} p(\v{k},t) \\
  &&=  i\frac{|d_{cv}|^2}{2\hbar\Gamma} \frac{2}{V}\!\! \sum_{\v{k}} (\rho^e_{E_t}\!+\!\rho^h_{E_t}\!-\!1) F(E_t,z,t).
\end{eqnarray*}

With a simple parabolic dispersion relation and converting the $\v{k}$-summation to a transverse energy integral, we obtain:
\begin{align}
\label{ptot_t_main}
P_{tot}(t) \!=\!  i\frac{|d_{cv}|^2}{2\hbar\Gamma} \!\int\! dE_t D_r^{2D} (\rho^e_{E_t}+\rho^h_{E_t}-1) F(E_t,z,t)
\end{align}
The dipole matrix element can be rewritten as the momentum matrix element via $|d_{cv}|^2 = \frac{q^2}{m_0^2 \omega_0^2} |\uv{e}\cdot \v{p}|^2$ where $q$ is the electron charge and $m_0$ the electron mass. The macroscopic polarization calculated in Eq.~(\ref{ptot_t_main}) serves as a source term for the forward and backward propagating electric fields in the laser.  The constants on the right hand side of Eq.~(\ref{tw_eqn}) can be combined to yield a gain coefficient 

\begin{align*}
g_0 &=  \frac{\Gamma_{xy}q^2 D^{2D}_r |\uv{e}_j \cdot \v{p}_{cv}|^2}{2 n_0 c \epsilon_0 m_0^2\Gamma}.
\end{align*}

To complete the derivation of the propagation equations, we include the effects of carrier gratings resulting from the interference between forward and backward waves. Our approach to modeling this spatial hole burning (SHB) is to follow the techniques of \cite{Homar1996}, \cite{Javaloyes2009} and \cite{Homar1996-2} and expand the QW population as follows
\begin{align}
\label{p_zexp}
\rho^{e,h}_{E_t} = \rho^{e,h}_{qw, E_t} + \rho_{g,E_t} e^{i2k_0z} + \rho^*_{g,E_t} e^{-i2k_0z} + ...
\end{align}
For simplicity, we have used a single variable for the carrier gratings for both electrons and holes. The filtered  field in the polarization also consists of forward and backward components:
\begin{align}
\label{F_zexp}
F &= F_+ e^{-ik_0 z} + F_- e^{ik_0 z}
\end{align}
Inserting Eqs.~(\ref{ptot_t_main})-(\ref{F_zexp}) in Eq.~(\ref{tw_eqn}) and keeping only the phase-matched terms we obtain the electric field equations:
\begin{align}
\label{wave_eq_simple}
\begin{split}
\pm\pd{E_\pm}{z}+ &\frac{1}{v_g}\! \pd{E_\pm}{t} \!=\! \frac{g_0}{2} \!\! \int \! \frac{dE_t}{\hbar\omega_0} (\rho^e_{qw, E_t}\!+\!\rho^h_{qw,E_t}\!-\!1) F_\pm(E_t, z,t)\\
 & + g_0 \int \frac{dE_t}{\hbar\omega_0} \rho^{(*)}_{g,E_t} F_\mp(E_t, z,t)
\end{split}
\end{align}
We note that the grating term $\rho_{g,E_t}^{(*)}$ is associated with the forward wave equation and its conjugate with the backward wave. Finally, we simply add the additional terms in Eq.~(\ref{wave_eq_simple}) that describe standard linear and nonlinear effects, and scale via $n_{qw}$, the number of quantum wells, to obtain                                                                                                                                 
\begin{align}
\begin{split}
\label{wave_eq}
\pm \pd{E_\pm}{z}+ &\frac{1}{v_g} \pd{E_\pm}{t} + i\frac{k''}{2} \pdd{E_\pm}{t} = \\
& -\frac{\alpha}{2} E_\pm - \left(\frac{\alpha_S}{2}+i\beta_S\right)(|E_\pm|^2+2|E_\mp|^2)E_\pm +S_{sp} \\
 & + n_{qw} \frac{g_0}{2} \int \frac{dE_t}{\hbar\omega_0} (\rho^e_{qw, E_t}+\rho^h_{qw,E_t}-1) F_\pm(E_t, z,t)\\
 & + n_{qw} g_0 \int \frac{dE_t}{\hbar\omega_0} \rho^{(*)}_{g,E_t} F_\mp(E_t, z,t)
\end{split}
\end{align}
where $k''$ is the dispersion coefficient, $\alpha$ is the linear waveguide loss, and $\alpha_S, \beta_S$ are respectively the two-photon absorption and Kerr nonlinear coefficients, and $S_{sp}$ is the spontaneous emission term derived in ref. \cite{MarkIEEE}.

For simplicity, we rewrite the complicated gain term as a function of $G_{\pm}(E_\pm)$ so that Eq.~(\ref{wave_eq}) is more compactly represented
\begin{align}
\begin{split}
\label{wave_eq_simple}
\pm \pd{E_\pm}{z}+ &\frac{1}{v_g} \pd{E_\pm}{t} + i\frac{k''}{2} \pdd{E_\pm}{t} = \\
& -\frac{\alpha}{2} E_\pm - \left(\frac{\alpha_S}{2}+i\beta_S\right)(|E_\pm|^2+2|E_\mp|^2)E_\pm \\
& +S_{sp}+G_{\pm}(E_\pm).
\end{split}
\end{align}
These field equations are coupled with the carrier rate equations for the SCH and QW sections, of which the complete forms are shown in the Appendix. 

We extend this gain model to the waveguide array structure with three waveguides. By coupling out
low-intensity components of the electric field to the neighboring waveguides, we can effectively shape the electric field propagating in the first waveguide through intensity discrimination, thus achieving highly robust stable waveform generation in the laser cavity.   The resulting approximate evolution dynamics describing the waveguide array mode-locking is thus given by
\begin{align}
\begin{split}
\label{wave_eq1}
\pm \pd{E^{1}_\pm}{z}+ &\frac{1}{v_g} \pd{E^{1}_\pm}{t} + i\frac{k''}{2} \pdd{E^{1}_\pm}{t} =  \\
 & -\frac{\alpha}{2} E^{1}_\pm - \left(\frac{\alpha_S}{2}+i\beta_S\right)(|E^{1}_\pm|^2+2|E^{1}_\mp|^2)E^{1}_\pm \\
 & +S_{sp} +G^{1}_{\pm}+iCE^{2}_\pm
\end{split}
\end{align}
\begin{align}
\begin{split}
\label{wave_eq2}
\pm \pd{E^{2}_\pm}{z}+ &\frac{1}{v_g} \pd{E^{2}_\pm}{t} + i\frac{k''}{2} \pdd{E^{2}_\pm}{t} =  \\
 & -\frac{\alpha}{2} E^{2}_\pm - \left(\frac{\alpha_S}{2}+i\beta_S\right)(|E^{2}_\pm|^2+2|E^{2}_\mp|^2)E^{2}_\pm \\
 & +S_{sp} +G^{2}_{\pm}+iC(E^{1}_\pm+E^{3}_\pm)
\end{split}
\end{align}
\begin{align}
\begin{split}
\label{wave_eq3}
\pm \pd{E^{3}_\pm}{z}+ &\frac{1}{v_g} \pd{E^{3}_\pm}{t} + i\frac{k''}{2} \pdd{E^{3}_\pm}{t} =  \\
 & -\frac{\alpha}{2} E^{3}_\pm - \left(\frac{\alpha_S}{2}+i\beta_S\right)(|E^{3}_\pm|^2+2|E^{3}_\mp|^2)E^{3}_\pm \\
 & +S_{sp} +G^{3}_{\pm} +iCE^{2}_\pm
\end{split}
\end{align}
Here the dimensionless coupling factor $C$ is determined by the design parameters of the WGA, such as the waveguide
separation. Thus it can be adjusted via designing the waveguide array to realize optimal mode-locking of the output. 


\section{Numerical Results for Stable Waveform generation}
\label{sec:results}

We solve the forward and backward wave equations (Eqs.~(\ref{wave_eq1})-(\ref{wave_eq3})), coupled with the carrier rate equations (Eqs. (\ref{sch_eq})-(\ref{pg_eq})) numerically using a robust predictor-corrector scheme which
generically improves stability properties compared to the Euler algorithm~\cite{Chang_numerical}.  We simulate $100$ ns of operation of the waveguide array starting from noise 
with a time step of $\Delta t = 30$ fs. The full simulation parameters are listed in Table~\ref{material_param} in Appendix C. Note that the major difference between GaAs and the previous material considered, InGaAsP, is that GaAs has a larger central transition frequency and diffusion coefficient while the two-photon absorption is significantly decreased. These changes of parameters prevent the stable waveform generation from the single waveguide model. Given the large parameter space to be explored using WGAs for optimal design, we focus on (i) the linear waveguide coupling coefficient $C$ between the three waveguides, (ii) the input pump to the waveguide array, and (iii) the depth of the quantum well. The other parameters are chosen to be the feasible parameters for an experimental design~\cite{MarkIEEE}.

Similar to the single waveguide model, we specify the limits of $E_t$ and the number of energy bins. 
Note that our simulations also suggested that higher energy carriers can contribute to the total gain thus affect the generation of the lasing dynamics.  We choose max$(E_t) = 50$ meV, with 25 energy bins to guarantee a small energy step for an accurate gain integral.


\begin{figure}[t]
\hspace*{.16in}
{\begin{overpic}
[width=0.95\linewidth]{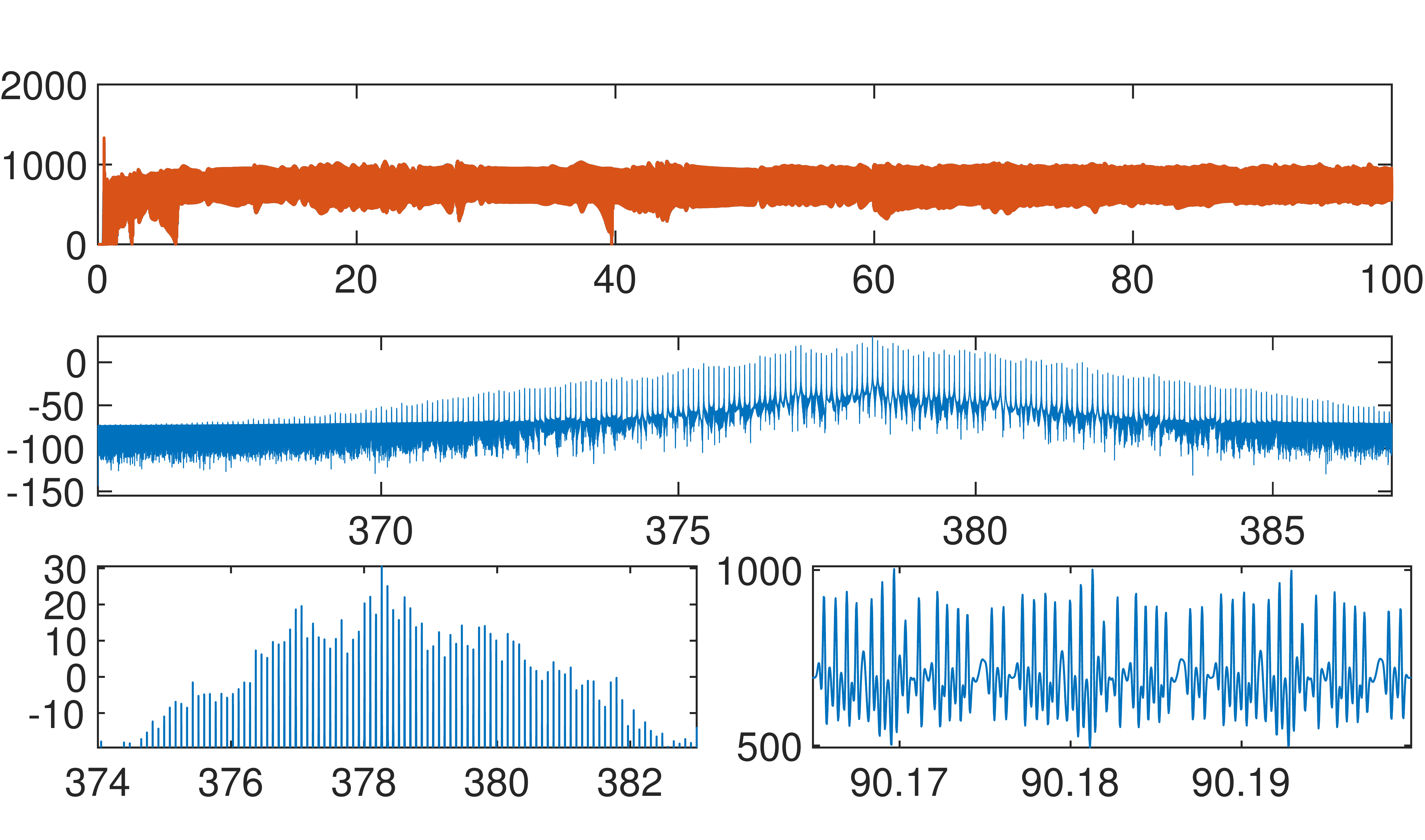}
\put(-5,49){{\small $P$}}
\put(84,37.5){\small $t$ (ns)}
\put(-5,29){{\small $|\hat{E}|^2$ }}
\put(73,20.5){\small $f$ (Hz)}
\put(50,14){{\small $P$}}
\put(73,-1){\small $t$ (ns)}
\put(-5,14){{\small $|\hat{E}|^2$ }}
\put(23,-1){\small $f$ (Hz)}
\put(10,49){\small (a)}
\put(10,31){\small (b)}
\put(10,15){\small (c)}
\put(59,15){\small (d)}
\end{overpic}}
\caption{Evolution of output power $P$ (mW) and power spectral density $|\hat{E}|^2$ (dBm/Hz).  (a) The temporal output of the first waveguide at $I_{in}$ = 100 mA with the coupling factor C=0. (b) The power spectral density of the temporal output in log scale. (c) (d) The zoomed power spectral density and temporal output of the first waveguide. The output electric field is in the chaotic form for every round trip and we found no sign of frequency combs.}
\label{C0}
\end{figure}


\begin{figure}[t]
\hspace*{.16in}
{\begin{overpic}
[width=0.95\linewidth]{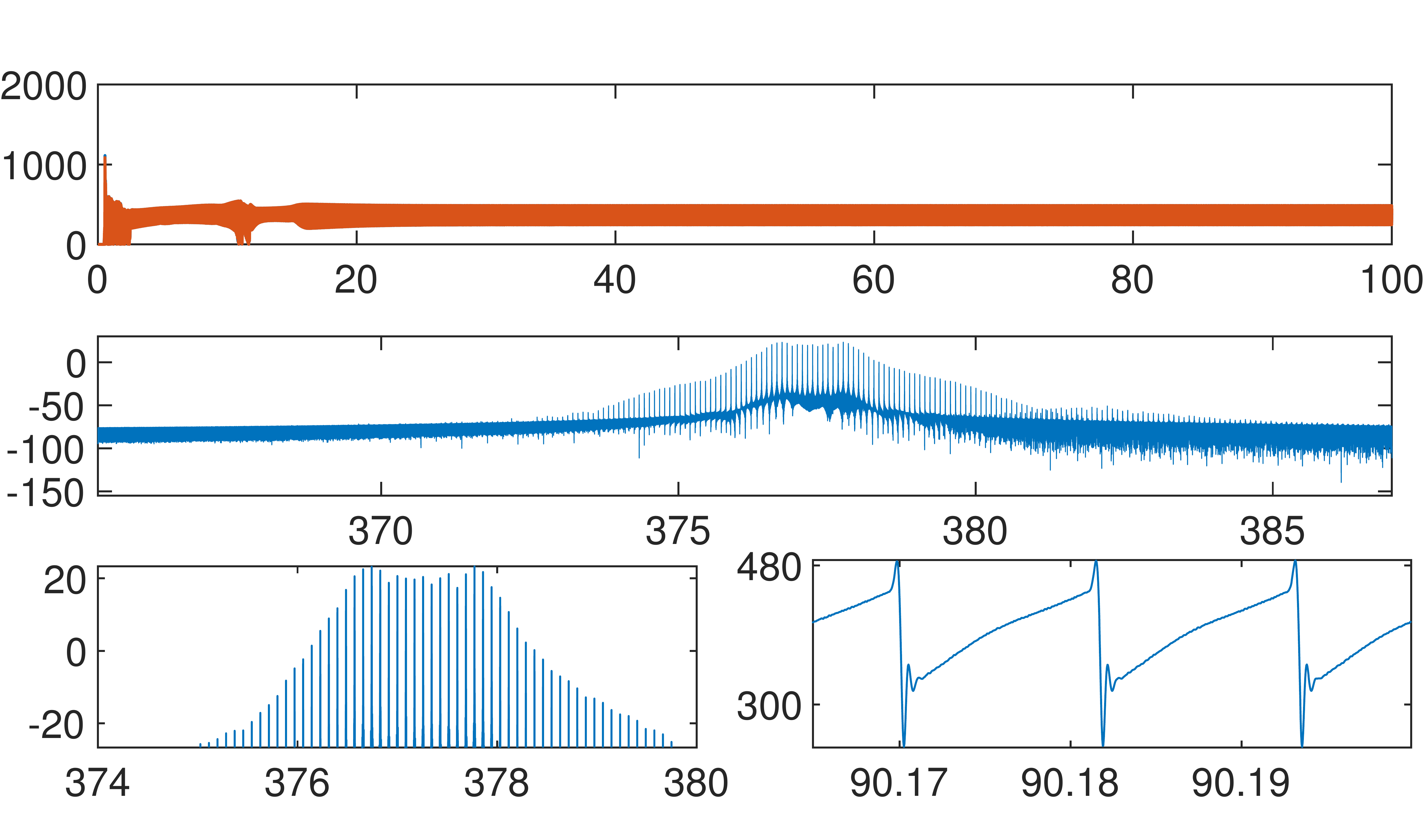}
\put(-5,49){{\small $P$}}
\put(84,37.5){\small $t$ (ns)}
\put(-5,29){{\small $|\hat{E}|^2$ }}
\put(73,20.5){\small $f$ (Hz)}
\put(50,14){{\small $P$}}
\put(73,-1){\small $t$ (ns)}
\put(-5,14){{\small $|\hat{E}|^2$ }}
\put(23,-1){\small $f$ (Hz)}
\put(10,49){\small (a)}
\put(10,31){\small (b)}
\put(10,15){\small (c)}
\put(59,15){\small (d)}
\end{overpic}}
\caption{Evolution of output power $P$ (mW) and power spectral density $|\hat{E}|^2$ (dBm/Hz). (a) The temporal output of the first waveguide at $I_{in}$ = 100 mA with the coupling factor C=1. A steady state is reached for $t>30$ ns. (b) The power spectral density of the temporal output in log scale shows a broad comb. (c) (d) The zoomed power spectral density and temporal output of the first waveguide. The output
is quasi-CW except for a short burst that repeats every round trip. Although it narrows down in the power spectral density, the spectrum is of higher intensity and a flatter top.}
\label{C1}
\end{figure}


\begin{figure}[htbp]
\hspace*{.16in}
{\begin{overpic}
[width=0.95\linewidth]{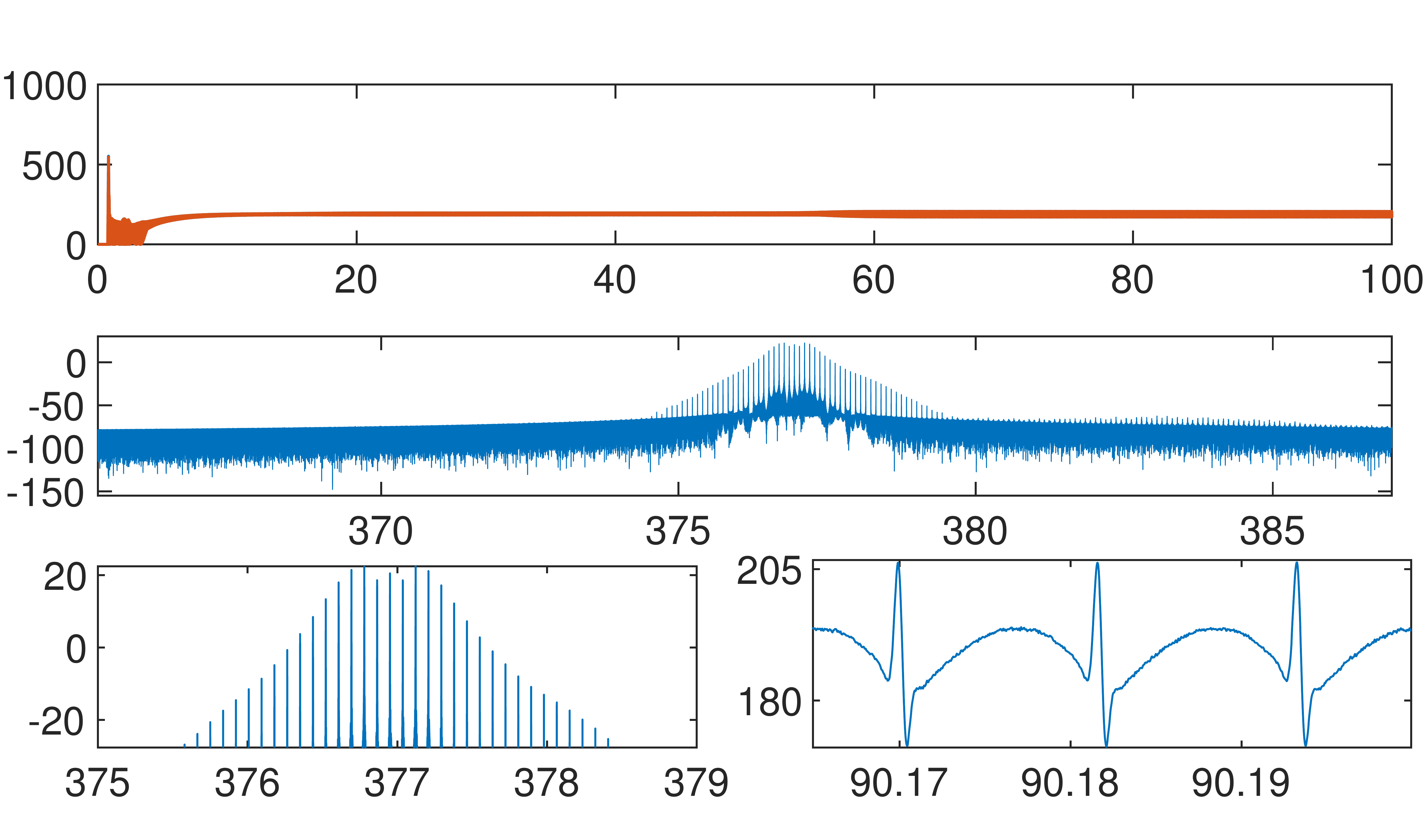}
\put(-5,49){{\small $P$}}
\put(84,37.5){\small $t$ (ns)}
\put(-5,29){{\small $|\hat{E}|^2$ }}
\put(73,20.5){\small $f$ (Hz)}
\put(50,14){{\small $P$}}
\put(73,-1){\small $t$ (ns)}
\put(-5,14){{\small $|\hat{E}|^2$ }}
\put(23,-1){\small $f$ (Hz)}
\put(10,49){\small (a)}
\put(10,31){\small (b)}
\put(10,15){\small (c)}
\put(59,15){\small (d)}
\end{overpic}}
\caption{Evolution of output power $P$ (mW) and power spectral density $|\hat{E}|^2$ (dBm/Hz). (a) The temporal output and (b) power spectral density of the temporal output in log scale of the first waveguide at $I_{in}$ = 100 mA with the coupling factor C=2. Longer time is needed to reach a steady state ($t>60$ ns). (c) (d) The zoomed power spectral density and temporal output of the first waveguide. The output power is further lowered and the spectral density is narrowed down as well.}
\label{C2}
\end{figure}

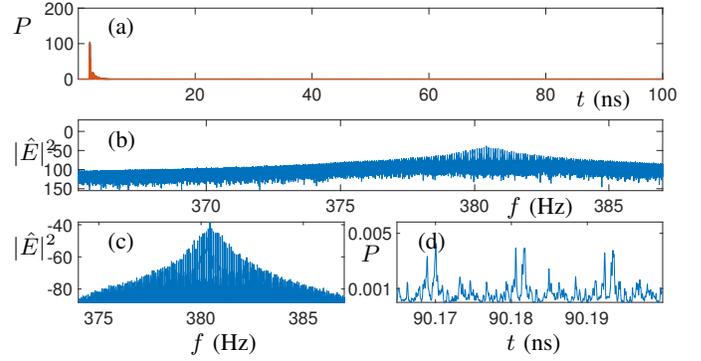
\begin{figure}[htbp]
\hspace*{.16in}
{\begin{overpic}
[width=0.95\linewidth]{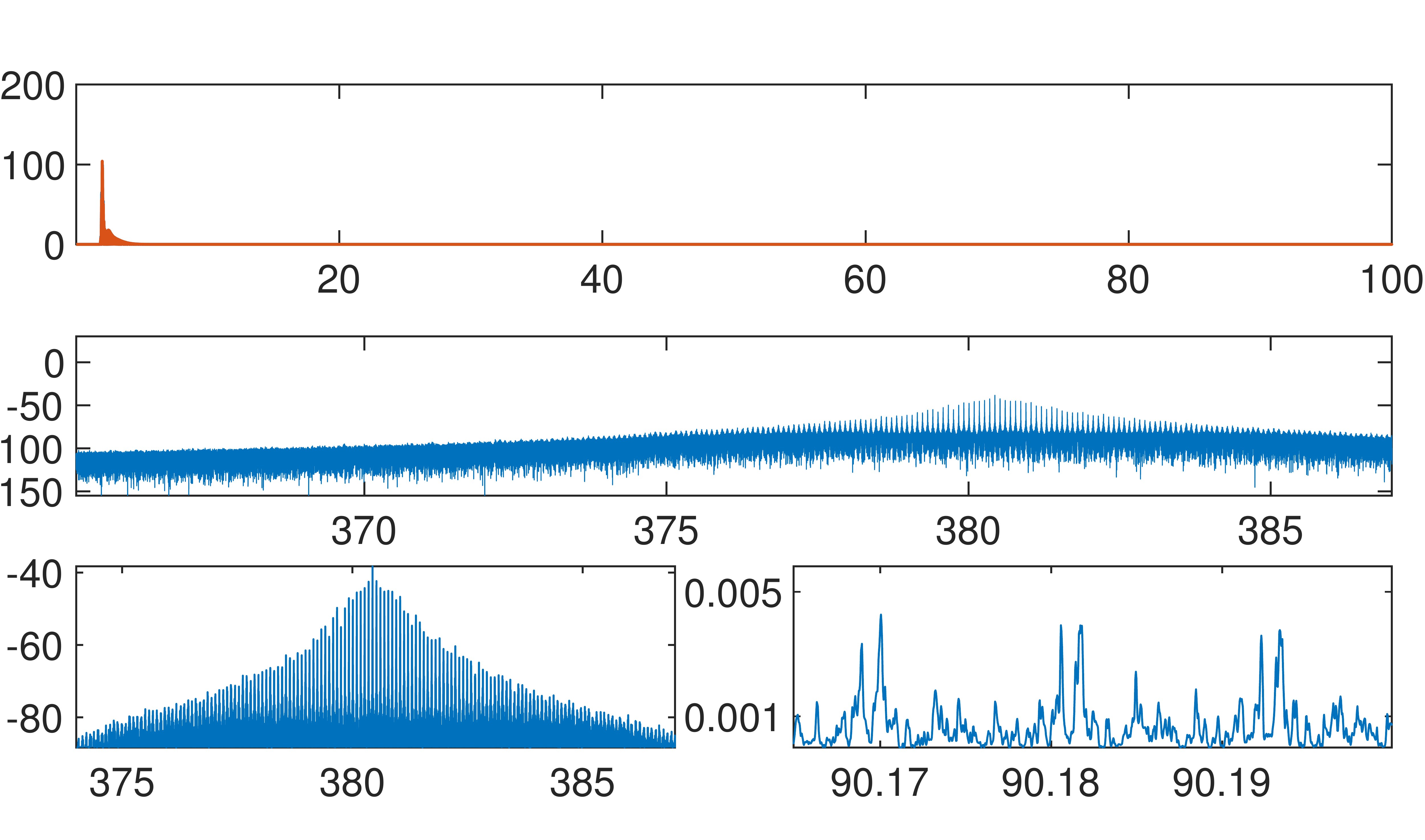}
\put(-5,49){{\small $P$}}
\put(84,37.5){\small $t$ (ns)}
\put(-5,29){{\small $|\hat{E}|^2$ }}
\put(73,20.5){\small $f$ (Hz)}
\put(50,13){{\small $P$}}
\put(73,-1){\small $t$ (ns)}
\put(-5,14){{\small $|\hat{E}|^2$ }}
\put(23,-1){\small $f$ (Hz)}
\put(10,49){\small (a)}
\put(10,31){\small (b)}
\put(10,15){\small (c)}
\put(59,15){\small (d)}
\end{overpic}}
\caption{Evolution of output power $P$ (mW) and power spectral density $|\hat{E}|^2$ (dBm/Hz). (a) The temporal output and (b) power spectral density of the temporal output in log scale of the first waveguide at $I_{in}$ = 100 mA with the coupling factor C=3. The output power quickly dies to a low value. (c) (d) The zoomed power spectral density and temporal output of the first waveguide. The gain is insufficient to pump the wave compared to the energy loss and we are left with the low-power white noises in the cavity.}
\label{C3}
\end{figure}


We first solved a single waveguide model by setting the coupling factor of $C=0$ so there is no coupling between waveguide. As shown in Fig. \ref{C0}, we are unable to generate the stable waveform since the electric field is highly variable. Turning on the coupling factor with a small value, e.g. $C=0.5$, does not stabilize the chaotic electric field in the cavity. When the coupling factor $C$ is increased to unity, the electric filed inside the cavity can be stabilized to its mode-locked periodic state quickly, see Fig. \ref{C1}. If we increase the value of the coupling factor $C$ from 1 to 2, the mode-locked state still holds in the cavity, but the output power of the electric field the first waveguide is decreased. This is reasonable since more energy gets coupled to the neighboring waveguides and dissipated in the 3rd waveguide. While the coupling factor gets too large, for example, $C=3$ in this case, a large amount of energy gets coupled to the neighboring waveguides and we are no longer able to get to an equilibrium, or balance, of the gain and the loss dynamics, thus the mode locked state disappears. The results are shown in Figs. \ref{C2}-\ref{C3}.

In the simulations above, the input current to the first waveguide is set to be $I_{in}$=100 mA. There is no energy pumping (gain from current injection) applied to the second and third waveguides in the array.  Waveguide two experiences a net intrinsic loss with $\alpha = 5$ cm$^{-1}$, whereas an extra electron and hole sweep out with lifetime $0.1$ ps and $0.3$ ps exists in waveguide three as it is reversely biased. 

With the energy step remaining as $2$ meV but max($E_{t}$) increased to 100 meV, the mode-locked state does not hold with the coupling factor $C=1$ and the electric filed inside waveguide one falls back to the chaotic state. Since the limit of $E_{t}$ equivalently contributes to the gain, we test another case when max($E_{t}$) remains 50 meV but the input pump on the first waveguide is increased to 200 mA.  Neither of these parameter regimes are capable of producing a repeatable waveform.

\begin{figure}[t]
\hspace*{.16in}
{\begin{overpic}
[width=0.95\linewidth]{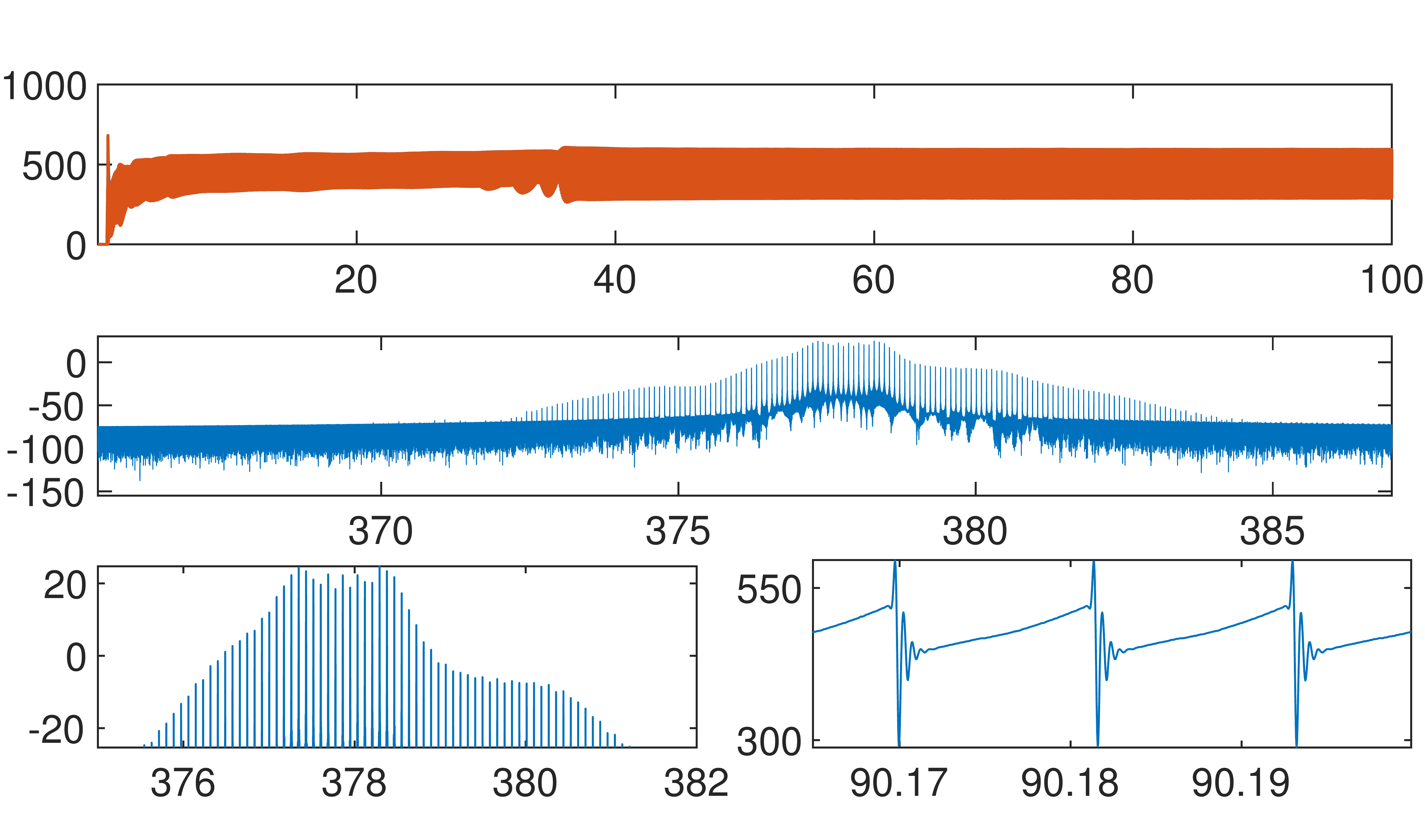}
\put(-5,49){{\small $P$}}
\put(84,37.5){\small $t$ (ns)}
\put(-5,29){{\small $|\hat{E}|^2$ }}
\put(73,20.5){\small $f$ (Hz)}
\put(50,13){{\small $P$}}
\put(73,-1){\small $t$ (ns)}
\put(-5,14){{\small $|\hat{E}|^2$ }}
\put(23,-1){\small $f$ (Hz)}
\put(10,49){\small (a)}
\put(10,31){\small (b)}
\put(10,15){\small (c)}
\put(59,15){\small (d)}
\end{overpic}}
\caption{Evolution of output power $P$ (mW) and power spectral density $|\hat{E}|^2$ (dBm/Hz). (a) The temporal output and (b) the power spectral density of the temporal output in log scale of the first waveguide at $I_{1in}$=100 mA and $I_{2in}$=30 mA with the coupling factor C=1. (c) (d) The zoomed power spectral density and temporal output of the first waveguide. Compared to Fig. \ref{C2},  the electric output still remains in a similar shape with a higher output power in the time domain.}
\label{C1_I100_I30}
\end{figure}

\begin{figure}[t]
\hspace*{.16in}
{\begin{overpic}
[width=0.95\linewidth]{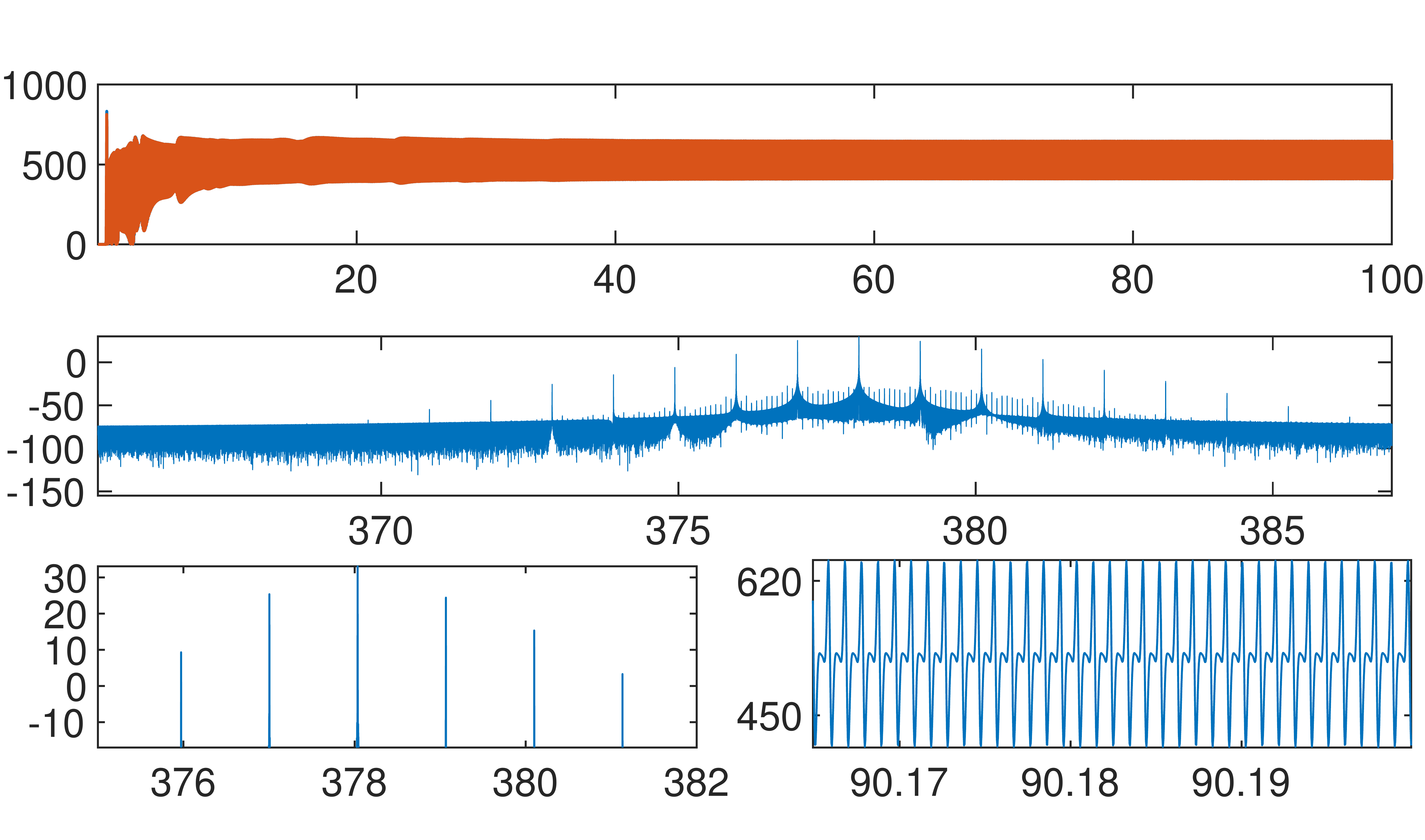}
\put(-5,49){{\small $P$}}
\put(84,37.5){\small $t$ (ns)}
\put(-5,29){{\small $|\hat{E}|^2$ }}
\put(73,20.5){\small $f$ (Hz)}
\put(50,13.5){{\small $P$}}
\put(73,-1){\small $t$ (ns)}
\put(-5,14){{\small $|\hat{E}|^2$ }}
\put(23,-1){\small $f$ (Hz)}
\put(10,49){\small (a)}
\put(10,31){\small (b)}
\put(10,15){\small (c)}
\put(59,15){\small (d)}
\end{overpic}}
\caption{Evolution of output power $P$ (mW) and power spectral density $|\hat{E}|^2$ (dBm/Hz). (a) The temporal output and (b) the power spectral density of the temporal output in log scale of the first waveguide at $I_{1in}$=100 mA and $I_{2in}$=50 mA with the coupling factor C=1. (c) (d) The zoomed power spectral density and temporal output of the first waveguide. The increased input pump to the second waveguide has a significant impact on the output of the waveguide array. The period of the electric output in the time domain is largely decreased while the power spectral density has a wider separation between each comb line.}
\label{C1_I100_I50}
\end{figure}

\begin{figure}[t]
\hspace*{.16in}
{\begin{overpic}
[width=0.95\linewidth]{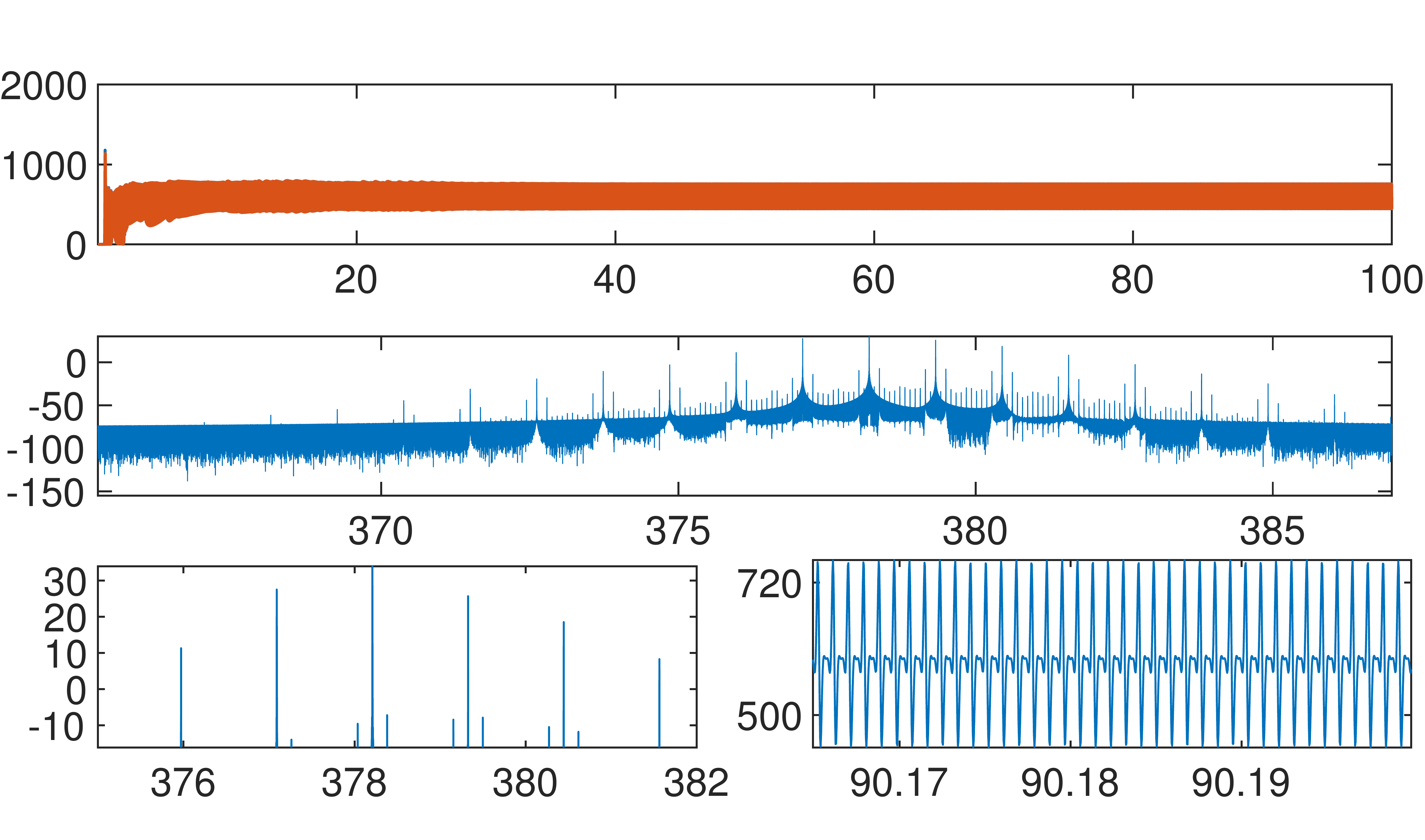}
\put(-5,49){{\small $P$}}
\put(84,37.5){\small $t$ (ns)}
\put(-5,29){{\small $|\hat{E}|^2$ }}
\put(73,20.5){\small $f$ (Hz)}
\put(50,13.5){{\small $P$}}
\put(73,-1){\small $t$ (ns)}
\put(-5,14){{\small $|\hat{E}|^2$ }}
\put(23,-1){\small $f$ (Hz)}
\put(10,49){\small (a)}
\put(10,31){\small (b)}
\put(10,15){\small (c)}
\put(59,15){\small (d)}
\end{overpic}}
\caption{Evolution of output power $P$ (mW) and power spectral density $|\hat{E}|^2$ (dBm/Hz). (a) The temporal output and (b) the power spectral density of the temporal output in log scale of the first waveguide at $I_{1in}$=100 mA and $I_{2in}$=80 mA with the coupling factor C=1. (c) (d) The zoomed power spectral density and temporal output of the first waveguide. Compared to Fig. \ref{C1_I100_I50}, extra comb lines arise around the central lines due to a higher input pump.}
\label{C1_I100_I80}
\end{figure}

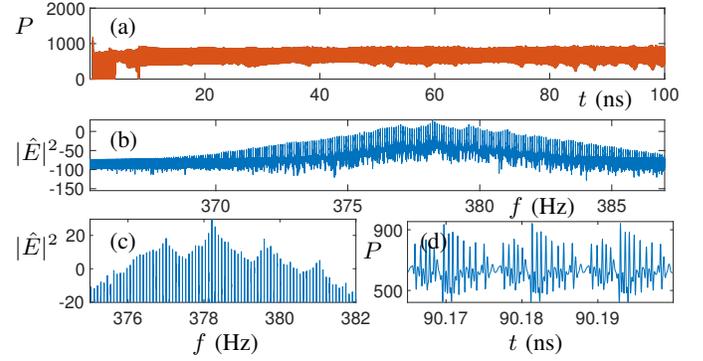
\begin{figure}[t]
\hspace*{.16in}
{\begin{overpic}
[width=0.95\linewidth]{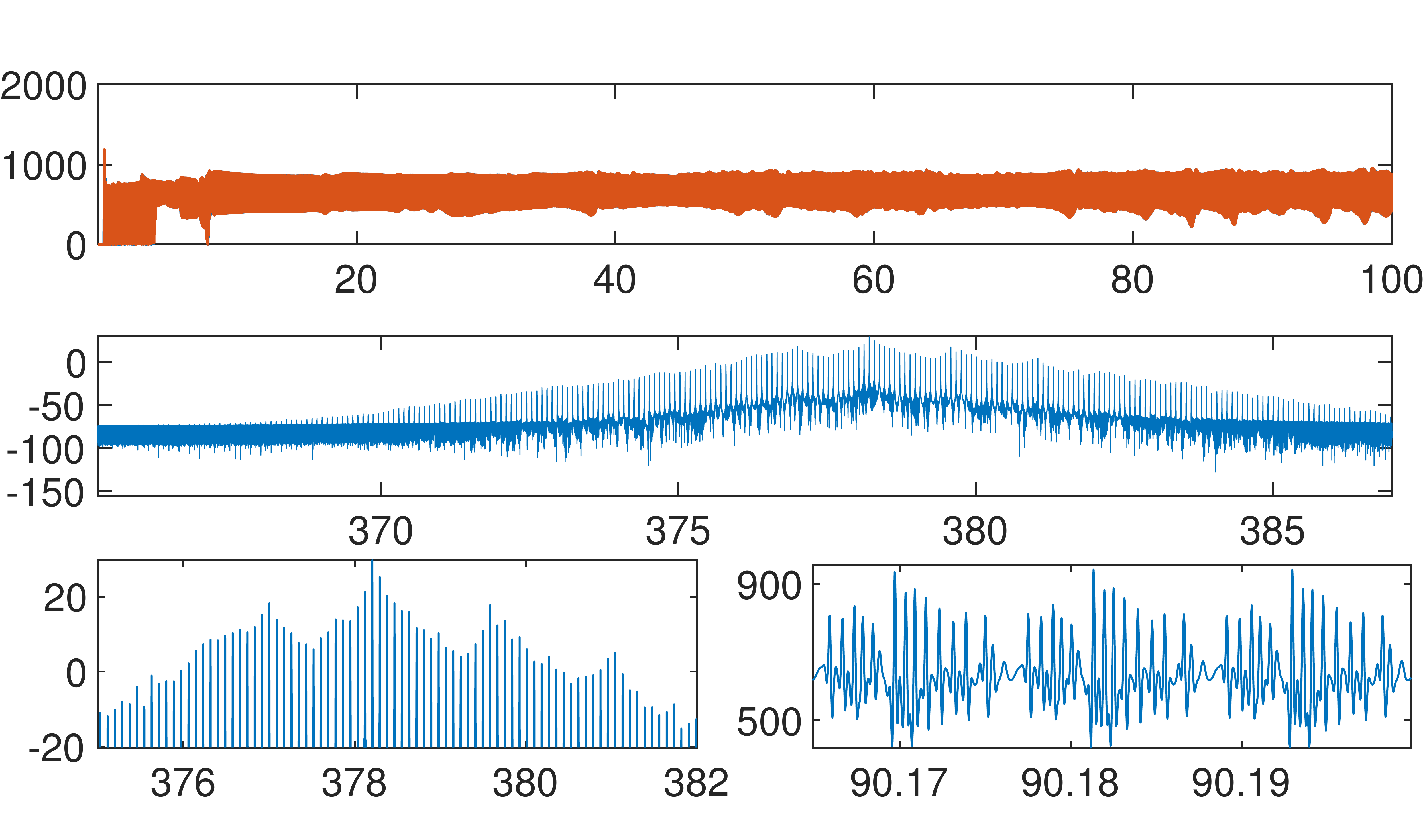}
\put(-5,49){{\small $P$}}
\put(84,37.5){\small $t$ (ns)}
\put(-5,29){{\small $|\hat{E}|^2$ }}
\put(73,20.5){\small $f$ (Hz)}
\put(50,13.5){{\small $P$}}
\put(73,-1){\small $t$ (ns)}
\put(-5,14){{\small $|\hat{E}|^2$ }}
\put(23,-1){\small $f$ (Hz)}
\put(10,49){\small (a)}
\put(10,31){\small (b)}
\put(10,15){\small (c)}
\put(59,15){\small (d)}
\end{overpic}}
\caption{Evolution of output power $P$ (mW) and power spectral density $|\hat{E}|^2$ (dBm/Hz). (a) The temporal output and (b) the power spectral density of the temporal output in log scale of the first waveguide at $I_{1in}$=100 mA and $I_{2in}$=100 mA with the coupling factor C=1. (c) (d) The zoomed power spectral density and temporal output of the first waveguide. The input pump to the second waveguide is sufficiently strong to break the balance between the gain and loss in the waveguide array, thus the mode-locked state is destroyed in this case.}
\label{C1_I100_I100}
\end{figure}

An extra gain to the second waveguide is added to make it nearly neutral from a gain-loss perspective in order to better shape the frequency combs. The results are shown in Fig. \ref{C1_I100_I30}-\ref{C1_I100_I100}, where the input current to the first waveguide is maintained as 100 mA and the coupling factor C is set to be 1. In contrast, the input current to the second waveguide is increased from 10 mA to 100 mA. Applying low input current, e.g. current less than 30 mA, to the second waveguide can slightly increase the output power but does not affect the results significantly, as shown in Fig. 
\ref{C1_I100_I30}. In contrast, when the input current is increased to 50 mA, the distance between each comb line is increased to about 1THz and the electric field oscillates in a shorter period in the time domain, perhaps indicating a harmonically modelocked state.  When the input current is increased to 80 mA, secondary comb lines appear around the central ones in the power spectral density but the electric field remains in the periodic state in the time domain. Increasing the input current to 100 mA destroys the balance between the loss and gain and the system falls into the chaotic state.

\section{Conclusion}
\label{sec:conclusion}
In conclusion, we have presented computational evidence that a traveling wave model for a quantum well and the mode-coupling in a waveguide array can generate frequency combs at 
800 nm. The mode coupling of the waveguide array provides the necessary intensity discrimination for pulse shaping stabilizing the generation of a repeatable waveform and frequency comb in the cavity. To experimentally realize stable, robust comblines, the coupling factor between waveguides in the array must be optimized. 

We explored the parameter space 
of WGA coupling factor $C$, input currents to the waveguides, waveguide biases, energy steps and energy limits, to understand the different performance characteristics of the waveguide array model and its dependency on the WGA parameters. The mode-locking behavior is sensitive to certain directions of the parameter space. Specifically, the increased input current, or depth of the quantum well, can break the balance between gain and loss, and the simulation shows destabilized comb generation with decreased hole capture time and numerical time step.  When also considering the spontaneous emission, the trends are consistent whereas the pulse shapes tend to vary slightly with the same parameters. Regardless, the numerical results demonstrate the generation of frequency combs at 
800 nm with the coupled waveguide array for a quantum well. This combined model of diode lasers can serve as an excellent candidate for compact, efficient and robust comb sources experimentally.

\appendix
\subsection{Derivation of the total polarization}
The material polarization $P_{tot}$ is obtained from the Bloch Equations as tailored to semiconductors \cite{ChowKochSargent1994}:

\begin{subequations}
\begin{align}
\begin{split}
\label{sc_bloch_p}
i \hbar \pd{p(\v{k},t)}{t} = (\hbar\omega_0 - \Delta E_{cv}(\v{k}) )p(\v{k},t) \\
- \frac{d_{cv} }{2}E(\v{k}, z,t)(\rho^e(\v{k},t)+ \rho^h(\v{k},t) - 1)-i\hbar \frac{p(\v{k}, t)}{T_2}
\end{split}
\end{align}
\begin{align}
\label{sc_bloch_ne}
\pd{\rho^e(\v{k},t)}{t} &= -\frac{1}{\hbar} Im[d^*_{cv} E^*(z,t) p(\v{k},t)] + \pd{\rho^e(\v{k},t)}{t}|_{relax}\\
\label{sc_bloch_nh}
\pd{\rho^h(\v{k},t)}{t} &= -\frac{1}{\hbar} Im[d^*_{cv} E^*(z,t) p(\v{k},t)] + \pd{\rho^h(\v{k},t)}{t}|_{relax}
\end{align}
\end{subequations}
where $p(\v{k},t)$ is the microscopic polarization, $\rho^{e,h}(\bf{k},t)$ is the occupation probability of electrons and holes, $d_{cv}$ is the dipole matrix element, $\Delta E_{cv}(\bf{k})$ is the transition energy between the conduction and valence bands, and $T_2 = 1/\Gamma$ is the intraband relaxation time which gives rise to homogenous broadening. It is important to note that these equations are in the time domain but are parameterized by the wavevector $\bf{k}$ and hence represent the time evolution of the subset of carriers with momentum $\bf{k}$.

A key simplification in our model is to assume that the intraband scattering is sufficiently fast to warrant the microscopic polarization adiabatically following the changes in carrier population. For modeling ultra-short pulses, this assumption may no longer hold and a full set of polarization equations will need to be solved dynamically. Integrating Eq.~(\ref{sc_bloch_p}), we obtain a time domain expression for the microscopic polarization in terms of the occupation probabilities and the electric field:

\begin{align}
\begin{split}
p(\v{k}, t) = \frac{i d_{cv} E(\v{k},z,t)}{2\hbar} \int_{-\infty}^t dt' E(z,t') \\
\times e^{ -\left(i\frac{\Delta E_{cv}(\v{k})}{\hbar}-\omega_0 \right)(t-t')-\Gamma(t-t')}(\rho^e(\v{k},t') + \rho^h(\v{k},t') - 1)
\end{split}
\end{align}
Next, we make the standard adiabatic approximation in which we assume the occupation probabilities evolve slowly compared to the intraband relaxation time $1/\Gamma$ and can be taken out of the integral, with $t'$ replaced by $t$. The remaining convolution integral is then defined as the filtered field \cite{Gioannini2015}

\begin{align}
\label{F_field}
F(\v{k}, z,t) = \Gamma \int_{-\infty}^t dt' e^{i(\frac{\Delta E_{cv}(\v{k}) }{\hbar}-\omega_0)(t-t')-\Gamma(t-t')} E(z,t')
\end{align}

The filtered field consists of all the components that interact with the population $\rho^{e,h}(\bf{k},t)$. Here the transition frequency is defined such that $\hbar\omega_0$ is the transition energy for a confined electron-hole pair with zero transverse energy and satisfies

\begin{align*}
\frac{\Delta E_{cv}(\v{k}) }{\hbar}-\omega_0 = \frac{E_t(\v{k})}{\hbar}
\end{align*}
Thus each discretized carrier group will have a different filtering frequency defined by the transverse energy $E_t$. The time-dependent microscopic polarization reduces to a simple expression:

\begin{align}
\label{p_ss}
p(\v{k}, t) = \frac{i d_{cv}}{2\hbar \Gamma} F(\v{k},z,t)
\end{align}
Here we note that physically, the $\bf{k}$ dependence of the confined carriers in the quantum well is due to a momentum $\bf{k}$ in the two transverse directions, and we therefore define a transverse energy with a simple parabolic band structure:

\begin{align}
E_t &= \frac{\hbar^2 |\v{k}|^2}{2 m_r^*}
\end{align}

\noindent where $m^*_r$ is the reduced effective mass. Hence to save space, we interchangeably write $\rho^{e,h}(\v{k},t) \leftrightarrow \rho^{e,h}_{E_t}$. We can also rewrite the filtered field by interchanging $F(\v{k},z,t) \leftrightarrow F(E_t,z,t)$.

The total polarization per volume is a summation over all carrier groups with momentum $\bf{k}$. Therefore, the total polarization for a 2-D quantum well can be written as:

\begin{align}
\begin{split}
\label{ptot_sumk}
P_{tot}(t) = \frac{2}{V} \sum_{\v{k}} d^*_{ev} p(\v{k},t) \\
= i\frac{|d_{cv}|^2}{2\hbar\Gamma} \frac{2}{V} \sum_{\v{k}} (\rho^e_{E_t}+\rho^h_{E_t}-1) F(E_t,z,t).
\end{split}
\end{align}

\noindent The $\v{k}$-summation can be converted to a transverse energy integral. We use a simple parabolic dispersion relation for the conduction and valence bands:

\begin{subequations}
\begin{align}
E_c &= E_g+ E_{e1}+\frac{\hbar^2 |\v{k}|^2}{2 m_e^*}\\
E_v &= E_{h1}-\frac{\hbar^2 |\v{k}|^2}{2 m_h^*}\\
\hbar\omega_0 &= E_g+E_{e1} -E_{h1}
\end{align}
\end{subequations}

\noindent where $E_g$ is the band gap energy, $E_{e1}$ is the confined electron energy, $E_{h1}$ is the confined hole energy, $m^*_{e,h}$ is the electron and hole effective mass (we have assumed only a single confined electron state). Rewriting Eq.~(\ref{ptot_sumk}) with an energy integral, we obtain:

\begin{align}
\label{ptot_t}
P_{tot}(t) =  i\frac{|d_{cv}|^2}{2\hbar\Gamma} \int dE_t D_r^{2D} (\rho^e_{E_t}+\rho^h_{E_t}-1) F(E_t,z,t)
\end{align}

\subsection{Carrier rate equations for the SCH and QW sections}
The QW equations are labeled with the transverse variable  for each discretized bin yielding

\begin{subequations}
\begin{align}
\begin{split}
\label{sch_eq}
\pd{\rho^{e,h}_{sch}}{t} = \frac{\eta J_{in}}{q N_{c,v,sch}h_{sch}}(1-\rho^{e,h}_{sch})- \frac{\rho^{e,h}_{sch}}{\tau_{sp}} \\
+n_{qw}\sum_{E_t} \left[\rho^{e,h}_{qw, E_t}\frac{(1-\rho^{e,h}_{sch})}{\tau^{e,h}_e}-\rho^{e,h}_{sch}\frac{(1-\rho^{e,h}_{qw,E_t})}{\tau^{e,h}_c}\right]
\end{split}
\end{align}
\begin{align}
\begin{split}
\label{qw_eq}
\pd{\rho^{e,h}_{qw, E_t}}{t} = \frac{h_{sch}N_{c,v,sch}}{n_{qw} h_{qw}N_{r,qw}}\left(\rho^{e,h}_{sch}\frac{(1-\rho^{e,h}_{qw,E_t})}{\tau^{e,h}_c} \right. \\
  \left. - \rho^{e,h}_{qw,E_t}\frac{(1-\rho^{e,h}_{sch})}{\tau^{e,h}_e}\right) 
- \frac{\rho^{e,h}_{qw, E_t}}{\tau_{sp}} - R_{st} - R_{g}
\end{split}
\end{align}
\begin{align}
\begin{split}
\label{pg_eq}
\pd{\rho_{g, E_t}}{t} &=- \frac{\rho_{g,E_t}}{\tau_{sp}} -4k_0^2 D \rho_{g,E_t} - 2g_0 \frac{\Delta E_t}{(\hbar\omega_0)^2 h_{qw} W N_{r,qw}}\\
& \times \left[ \frac{1}{2}(E_+^*F_- + F_+^* E_-)(\rho^e_{qw}+\rho^h_{qw}-1) \right. \\
& \left. + 2\text{Re}(E^*_+F_+ + E^*_-F_-)\rho_{g,E_t}\right]
\end{split}
\end{align}
\end{subequations}

\begin{align}
R_{st} &= 2g_0 \frac{\Delta E_t}{(\hbar\omega_0)^2 h_{qw} W N_{r,qw}} (\rho_{qw, E_t}^e+\rho^h_{qw, E_t}-1) \text{Re}(E^* F)
\end{align}

\begin{align}
\begin{split}
R_g &= 2g_0 \frac{\Delta E_t}{(\hbar\omega_0)^2 h_{qw} W N_{r,qw}} \big((E_+F^*_- + F_+ E^*_-)\rho_{g,E_t} \\
& + (E_+^*F_- + F_+^* E_-)\rho^*_{g,E_t}\big)
\end{split}
\end{align}

\noindent where $N_{c,v,sch} = 2 \left(\frac{m_{e,h}^* k_B T}{2\hbar^2 \pi}\right)^{3/2}$, $N_r = \frac{m_{r}^* \Delta E_t }{\hbar^2 \pi h_{qw}}$ are the effective 3-D and 2-D density of states, $D$ is the ambipolar diffusion coefficient, $\tau_{sp}$ is the spontaneous emission lifetime, $\tau^{e,h}_c$ is the capture lifetime, and $\tau^{e,h}_e$ is the escape lifetime. The recombination rates $R_{st}$ and $R_g$ govern population decay due to stimulated emission and the carrier grating respectively. The escape times $\tau_e^{e,h}$ are particularly important in our model as they phenomenologically represent intraband interactions.  As shown in the Appendix, they are given by 

\begin{align}
\tau^e_e = \tau^e_c \exp( (\delta E_c-\frac{m^*_r}{m^*_e}E_t)/k_B T)\\
\tau^h_e = \tau^h_c \exp( (\delta E_v-\frac{m^*_r}{m^*_h}E_t)/k_B T)
\end{align}
The value of these escape times is tailored specifically to allow the rate equations \ref{sch_eq}, \ref{qw_eq} to relax to the Fermi-Dirac distribution.

\subsection{Simulation parameters for the GaAs system}

Table~\ref{ta:para} shows the various parameters used in the simulation.  Most of the parameters are extracted directly from previous physically realizable configurations and materials.

\begin{table}[t]
\centering
\caption{\bf Simulation parameters for the GaAs system. \label{ta:para}}
\begin{tabular}{lllllll}
\hline
{\small Symbol} & Description & Value\\ \hline
$L$ & Length of device & 500 $\mu$m \\ \hline
$W$ & Width of waveguide & 4 $\mu$m \\ \hline
$h_{sch}$ & Height of SCH layer & 50 nm \\ \hline
$h_{qw}$ & Height of quantum well & 5 nm \\ \hline
$n_0$ & Group refractive index & 3.5 \\ \hline
$n_{qw}$ & Number of quantum wells & 2 \\ \hline
$\alpha$ & Intrinsic waveguide loss & 5 cm$^{-1}$ \\ \hline
$\Gamma_{xy}$ & Optical confinement factor & 0.02 \\ \hline
$\alpha_S$ & Two-photon absorption & 580 W$^{-1}$m$^{-1}$\\ \hline
$\beta_S$ & Kerr coefficient &  430 W$^{-1}$m$^{-1}$\\ \hline
$\hbar \omega_0$ & Central transition energy & 1.55 eV\\ \hline
$|\uv{e}\cdot\v{p}|^2 $ & Momentum matrix element & 25 meV$\!\times \! m_0/6$ \\ \hline
$\Gamma$ & Homogenous half linewidth & 11 meV/$\hbar$ \\ \hline
$m^*_{e, sch}$ & {\small Effective electron mass in SCH layer} & $0.125 m_0$ \\ \hline
$m^*_{h, sch}$ &  {\small Effective mass of holes in SCH layer} & $0.703 m_0$ \\ \hline
$m^*_{e, qw}$ &  {\small Effective electron mass in GaAs QW} & $0.093 m_0$\\ \hline
$m^*_{h, qw}$ &  {\small Effective mass of holes in GaAs QW} & $0.53 m_0$ \\ \hline
$\tau_c^{e, h,qw}$ & electron, hole capture time  & $1$, 10 ps \\ \hline
$\delta E_c$ &  {\small Conduction band quantum well barrier} & 50 meV\\ \hline
$\delta E_v$ & Valence band quantum well barrier &  25 meV\\ \hline
$\beta_{sp}$ &  {\small Spontaneous emission coupling factor}& $1\times 10^{-4}$ \\ \hline
$\tau_{sp}$ & Spontaneous emission lifetime & 1 ns \\ \hline
$D$ & Ambipolar diffusion coefficient & 20 cm$^2$/s \\ \hline
\end{tabular}
  \label{material_param}
\end{table}

\section*{Acknowledgment}

J. N. Kutz acknowledges support from the Air Force Office of Scientific Research (FA9550-17-1-0200).
This research was also developed in part with funding from the Defense Advanced Research Projects Agency (DARPA) through the SCOUT program. The views, opinions and/or findings expressed are those of the author and should not be interpreted as representing the official views or policies of the Department of Defense or the U.S. Government. This research was also supported in part through computational resources and services provided by Advanced Research Computing at the University of Michigan, Ann Arbor.

\bibliographystyle{IEEEtran}  
\bibliography{QWLaser_Master}

\providecommand{\noopsort}[1]{}\providecommand{\singleletter}[1]{#1}%
\begin{thebibliography}{10}
\providecommand{\url}[1]{#1}
\csname url@samestyle\endcsname
\providecommand{\newblock}{\relax}
\providecommand{\bibinfo}[2]{#2}
\providecommand{\BIBentrySTDinterwordspacing}{\spaceskip=0pt\relax}
\providecommand{\BIBentryALTinterwordstretchfactor}{4}
\providecommand{\BIBentryALTinterwordspacing}{\spaceskip=\fontdimen2\font plus
\BIBentryALTinterwordstretchfactor\fontdimen3\font minus
  \fontdimen4\font\relax}
\providecommand{\BIBforeignlanguage}[2]{{%
\expandafter\ifx\csname l@#1\endcsname\relax
\typeout{** WARNING: IEEEtran.bst: No hyphenation pattern has been}%
\typeout{** loaded for the language `#1'. Using the pattern for}%
\typeout{** the default language instead.}%
\else
\language=\csname l@#1\endcsname
\fi
#2}}
\providecommand{\BIBdecl}{\relax}
\BIBdecl

\bibitem{Cundiff2003}
S.~T. Cundiff and J.~Ye, ``Femtosecond optical frequency combs,'' \emph{Rev. of
  Mod. Phys.}, vol.~75, no.~1, p. 325, March 2003.

\bibitem{Udem1999}
T.~Udem, J.~Reichert, R.~Holzwarth, and T.~W. H{\"a}nsch, ``Absolute optical
  frequency measurement of the cesium d1 line with a mode-locked laser,''
  \emph{Phys. Rev. Lett.}, vol.~82, no.~18, p. 3568, 1999.

\bibitem{Coddington2008}
I.~Coddington, W.~C. Swann, and N.~R. Newbury, ``Coherent multiheterodyne
  spectroscopy using stabilized optical frequency combs,'' \emph{Phys. Rev.
  Lett.}, vol. 100, p. 13902, 2008.

\bibitem{Diddams2001}
S.~A. Diddams, T.~Udem, J.~C. Bergquist, E.~A. Curtis, R.~E. Drullinger,
  L.~Hollberg, W.~M. Itano, W.~D. Lee, C.~W. Oates, K.~R. Vogel, and D.~J.
  Wineland, ``An optical clock based on a single trapped 199 hg1+ ion,''
  \emph{Science}, vol. 293, p. 825, 2001.

\bibitem{Cundiff2010}
S.~T. Cundiff and A.~M. Weiner, ``Optical arbitrary waveform generation,''
  \emph{Nat. Phot.}, vol.~4, p. 760, 2010.

\bibitem{Delfyett1992}
P.~J. Delfyett, L.~T. Florez, N.~Stoffel, T.~Gmitter, N.~C. Andreadakis,
  Y.~Silberberg, and J.~P. Heritage, ``High-power ultrafast laser diodes,''
  \emph{IEEE J. Quantum Electron.}, vol.~28, no.~10, p. 2203, 1992.

\bibitem{Tiemeijer1989}
L.~F. Tiemeijer, P.~I. Kuindersma, P.~J.~A. Thijs, and G.~L.~J. Rikken,
  ``Passive fm locking in ingaasp semiconductor lasers,'' \emph{IEEE J. Quantum
  Electron.}, vol.~25, no.~6, p. 1385, 1989.

\bibitem{MarkIEEE}
M.~Dong, N.~M. Mangan, J.~N. Kutz, S.~T. Cundiff, and H.~G. Winful, ``Traveling
  wave model for frequency comb generation in single-section quantum well diode
  lasers,'' \emph{IEEE Journal of Quantum Electronics}, vol.~53, no.~6, pp.
  1--11, 2017.

\bibitem{WilliamsWGA}
X.~Zhang, M.~Williams, S.~T. Cundiff, and J.~N. Kutz, ``Semiconductor diode
  laser mode-locking by a waveguide array,'' \emph{IEEE Journal of Selected
  Topics in Quantum Electronics}, vol.~22, no.~2, pp. 34--39, 2016.

\bibitem{Sutter1999}
D.~H. Sutter, G.~Steinmeyer, L.~Gallmann, N.~Matuschek, F.~Morier-Genoud,
  U.~Keller, V.~Scheuer, G.~Angelow, and T.~Tschudi, ``Semiconductor
  saturable-absorber mirror- assisted kerr-lens mode-locked ti:sapphire laser
  producing pulses in the two-cycle regime,'' \emph{Opt. Lett.}, vol.~24, p.
  631, 1999.

\bibitem{Fermann2013}
M.~E. Fermann and I.~Hartl, ``Ultrafast fibre lasers,'' \emph{Nat. Phot.},
  vol.~7, p. 868, 2013.

\bibitem{Herr2012}
T.~Herr, K.~Hartinger, J.~Riemensberger, C.~Y. Wang, E.~Gavartin, R.~Holzwarth,
  M.~L. Gorodetsky, and T.~J. Kippenberg, ``Universal formation dynamics and
  noise of kerr-frequency combs in microresonators,'' \emph{Nat. Phot.},
  vol.~6, p. 480, 2012.

\bibitem{Moskalenko2017}
V.~Moskalenko, J.~Koelemeij, K.~Williams, and E.~Bente, ``Study of extra wide
  coherent optical combs generated by a qw-based integrated passively
  mode-locked ring laser,'' \emph{Opt. Letters}, vol.~42, no.~7, p. 1428, 2017.

\bibitem{Rosales2011}
R.~Rosales, K.~Merghem, A.~Martinez, A.~Akrout, J.-P. Tourrenc, A.~Accard,
  F.~Lelarge, and A.~Ramdane, ``Inas/inp quantum-dot passively mode-locked
  lasers for 1.55-$\mu$m applications,'' \emph{IEEE J. Sel. Top. Quantum
  Electron.}, vol.~17, no.~5, p. 1292, 2011.

\bibitem{Gioannini2015}
M.~Gioannini, P.~Bardella, and I.~Montrosset, ``Time-domain traveling-wave
  analysis of the multimode dynamics of quantum dot fabry--perot lasers,''
  \emph{IEEE Sel. Topics Quantum Electron.}, vol.~21, no.~6, p. 1900811,
  December 2015.

\bibitem{Rosales2012}
R.~Rosales, K.~Merghem, C.~Calo, G.~Bouwmans, I.~Krestnikov, A.~Martinez, and
  A.~Ramdane, ``Optical pulse generation in single section inas/gaas quantum
  dot edge emitting lasers under continuous wave operation,'' \emph{App. Phys.
  Lett.}, vol. 101, p. 221113, 2012.

\bibitem{Rosales2012-2}
R.~Rosales, S.~G. Murdoch, R.~Watts, K.~Merghem, A.~Martinez, F.~Lelarge,
  A.~Accard, L.~P. Barry, and A.~Ramdane, ``High performance mode locking
  characteristics of single section quantum dash lasers,'' \emph{Optics
  Express}, vol.~20, no.~8, p. 8649, 2012.

\bibitem{Sato2003}
K.~Sato, ``Optical pulse generation using fabry--p{\'e}rot lasers under
  continuous-wave operation,'' \emph{IEEE J. Sel. Top. Quantum Electron.},
  vol.~9, no.~5, p. 1288, 2003.

\bibitem{Calo2015}
C.~Cal{\`o}, V.~Vujicic, R.~Watts, C.~Browning, K.~Merghem, V.~Panapakkam,
  F.~Lelarge, A.~Martinez, B.-E. Benkelfat, A.~Ramdane, and L.~P. Barry,
  ``Single-section quantum well mode-locked laser for 400 gb/s ssb-ofdm
  transmission,'' \emph{Opt. Express}, vol.~23, no.~20, p. 26442, 2015.

\bibitem{Homar1996}
M.~Homar, S.~Balle, and M.~S. Miguel, ``Mode competition in a fabry-perot
  semiconductor laser: travelling wave model with asymmetric dynamical gain,''
  \emph{Optics Communications}, vol. 131, pp. 380--390, 1996.

\bibitem{Arakawa1986}
Y.~Arakawa and A.~Yariv, ``Quantum well lasers - gain, spectra, dynamics,''
  \emph{IEEE J. Quantum Electron.}, vol. QE22, no.~9, p. 1887, 1986.

\bibitem{KN2010}
S.~N. Kaunga-Nyirenda, M.~P. Dlubek, A.~J. Phillips, J.~J. Lim, E.~C. Larkins,
  and S.~Sujecki, ``Theoretical investigation of the role of optically induced
  carrier pulsations in wave mixing in semiconductor optical amplifiers,''
  \emph{J. Opt. Soc. Am. B}, vol.~27, no.~2, p. 168, February 2010.

\bibitem{McDonald1995}
D.~McDonald and R.~F. O'Dowd, ``Comparison of two- and three-level rate
  equations in the modeling of quantum-well lasers,'' \emph{IEEE J. Quantum
  Electron.}, vol.~31, no.~11, p. 1927, November 1995.

\bibitem{Jones1995}
D.~J. Jones, L.~M. Zhang, J.~E. Carroll, and D.~D. Marcenac, ``Dynamics of
  monolithic passively mode-locked semiconductor lasers,'' \emph{IEEE J.
  Quantum Electron.}, vol.~31, no.~6, pp. 1051--1058, June 1995.

\bibitem{Vandermeer2005}
A.~D. Vandermeer and D.~T. Cassidy, ``A rate equation model of asymmetric
  multiple quantum-well lasers,'' \emph{IEEE J. Quantum Electron.}, vol.~41,
  no.~7, p. 917, 2005.

\bibitem{Gordon2008}
A.~Gordon, C.~Y. Wang, L.~Diehl, F.~X. K{\"a}rtner, A.~Belyanin, D.~Bour,
  S.~Corzine, G.~H{\"o}fler, H.~C. Liu, H.~Schneider, T.~Maier, M.~Troccoli,
  J.~Faist, and F.~Capasso, ``Multimode regimes in quantum cascade lasers: From
  coherent instabilities to spatial hole burning,'' \emph{Phys. Rev. A},
  vol.~77, no.~5, p. 053804, 2008.

\bibitem{Lenstra2014}
D.~Lenstra and M.~Yousefi, ``Rate-equation model for multi-mode semiconductor
  lasers with spatial hole burning,'' \emph{Opt. Express}, vol.~22, no.~7, p.
  8144, 2014.

\bibitem{Chow2002}
W.~W. Chow, H.~C. Schneider, S.~W. Koch, C.-H. Chang, L.~Chrostowski, and C.~J.
  Chang-Hasnain, ``Nonequilibrium model for semiconductor laser modulation
  response,'' \emph{IEEE J. Quantum Electron.}, vol.~38, no.~4, p. 402, 2002.

\bibitem{Proctor05}
J.~L. Proctor and J.~N. Kutz, ``Passive mode-locking by use of waveguide
  arrays,'' \emph{Optics letters}, vol.~30, no.~15, pp. 2013--2015, 2005.

\bibitem{Kutz08}
J.~N. Kutz and B.~Sandstede, ``Theory of passive harmonic mode-locking using
  waveguide arrays,'' \emph{Optics Express}, vol.~16, no.~2, pp. 636--650,
  2008.

\bibitem{Ching12}
Q.~Chao, D.~D. Hudson, J.~N. Kutz, and S.~Cundiff, ``Waveguide array fiber
  laser,'' \emph{IEEE Photonics Journal}, vol.~4, no.~5, pp. 1438--1442, 2012.

\bibitem{winful1}
L.~Rahman and H.~G. Winful, ``Nonlinear dynamics of semiconductor laser arrays:
  a mean field model,'' \emph{IEEE journal of quantum electronics}, vol.~30,
  no.~6, pp. 1405--1416, 1994.

\bibitem{winful2}
------, ``Improved coupled-mode theory for the dynamics of semiconductor laser
  arrays,'' \emph{Optics letters}, vol.~18, no.~2, pp. 128--130, 1993.

\bibitem{christo}
D.~Christodoulides and R.~Joseph, ``Discrete self-focusing in nonlinear arrays
  of coupled waveguides,'' \emph{Optics letters}, vol.~13, no.~9, pp. 794--796,
  1988.

\bibitem{roberto}
U.~Peschel, R.~Morandotti, J.~M. Arnold, J.~S. Aitchison, H.~S. Eisenberg,
  Y.~Silberberg, T.~Pertsch, and F.~Lederer, ``Optical discrete solitons in
  waveguide arrays. 2. dynamic properties,'' \emph{JOSA B}, vol.~19, no.~11,
  pp. 2637--2644, 2002.

\bibitem{jstqe09_wga}
B.~G. Bale, J.~N. Kutz, and B.~Sandstede, ``Optimizing waveguide array
  mode-locking for high-power fiber lasers,'' \emph{IEEE Journal of Selected
  Topics in Quantum Electronics}, vol.~15, no.~1, pp. 220--231, 2009.

\bibitem{Javaloyes2009}
J.~Javaloyes and S.~Balle, ``Emission directionality of semiconductor ring
  lasers: A traveling-wave description,'' \emph{IEEE J. Quantum Electron.},
  vol.~45, no.~5, p. 431, 2009.

\bibitem{Homar1996-2}
M.~Homar, J.~V. Moloney, and M.~S. Miguel, ``Traveling wave model of a
  multimode fabry-perot laser in free running and external cavity
  configurations,'' \emph{IEEE J. Quantum Electron.}, vol.~32, no.~3, p. 553,
  1996.

\bibitem{Chang_numerical}
C.~Sun, N.~Mangan, M.~Dong, H.~Winful, S.~Cundiff, and J.~N. Kutz, ``Stable
  numerical schemes for nonlinear dispersive equations with counter-propagation
  and gain dynamics,'' \emph{arxiv 2019}.

\bibitem{ChowKochSargent1994}
W.~W. Chow, S.~W. Koch, and M.~S. III, \emph{Semiconductor-Laser
  Physics}.\hskip 1em plus 0.5em minus 0.4em\relax Springer-Verlag, 1994.

\end{thebibliography}

\begin{IEEEbiographynophoto}{Chang Sun}
Chang Sun received the B.S. degree in applied physics
from University of Science and Technology of China, Hefei, Anhui, China, in 2015,
and she is currently pursuing the Ph.D.
degree of physics in University of Washington, Seattle, WA, USA.
\end{IEEEbiographynophoto}

\begin{IEEEbiographynophoto}{Mark Dong}
Mark Dong (M'17) received the B.S. degree in applied engineering physics
and electrical engineering from Cornell University, Ithaca, NY, USA, in 2012,
and the Ph.D. degree in electrical engineering from the University of Michigan,
Ann Arbor, MI, USA, in 2018.
\end{IEEEbiographynophoto}

\begin{IEEEbiographynophoto}{Niall M. Mangan}
Niall M. Mangan received the Dual B.S. degrees
in mathematics and physics, with a minor in chemistry,
from the Clarkson University, Potsdam, NY,
USA, in 2008, and the Ph.D. degree in systems
biology from Harvard University, Cambridge, MA,
USA, in 2013. She is currently an Assistant Professor of
engineering sciences and applied mathematics with
Northwestern University.
\end{IEEEbiographynophoto}

\begin{IEEEbiographynophoto}{Steven T. Cundiff}
Steven T. Cundiff received the B.A. degree in physics from Rutgers
University in 1985, and the M.S and Ph.D. degrees in applied physics from the
University of Michigan in 1991 and 1992, respectively. From 1993 to 1994,
he was a Post-Doctoral Scientist with the University of Marburg, Germany,
and from 1995 to 1997, he was a Post-Doctoral Member of the Technical
Staff at Bell Laboratories, Holmdel, NJ, USA. In 1997, he joined JILA, a joint
institute in Boulder, Co, USA, between the National Institute of Standards and
Technology and the University of Colorado, where he was a JILA Fellow and
a Professor Adjoint. From 2004 to 2009, he served as the Chief of the NIST
Quantum Physics Division. In 2015, he joined the Faculty at the University
of Michigan, where he is currently the Harrison M. Randall Collegiate
Professor of physics and a Professor of electrical and computer engineering.
He is a fellow of the Optical Society of America (OSA), the American
Physical Society, the Institute of Electrical and Electronics Engineers, and
the American Association for the Advancement of Science. He received the
Meggers Award from OSA and a Research Award from the Alexander von
Humboldt Foundation.
\end{IEEEbiographynophoto}

\begin{IEEEbiographynophoto}{Herbert G. Winful}
Herbert G. Winful (F'94) was born in London,
U.K., in 1952. He received the B.S. degree in electrical
engineering from the Massachusetts Institute
of Technology, Cambridge, MA, USA, in 1975, and
the Ph.D. degree in electrical engineering from the
University of Southern California, Los Angeles, CA,
USA, in 1981.
From 1980 to 1986, he was a member of the
Technical Staff at GTE Laboratories, Waltham, MA,
USA, where he conducted research in nonlinear
optics, semiconductor laser physics, and fiber optics.
He joined the EECS Department, University of Michigan, as an Associate
Professor in 1987 and was promoted to Full Professor in 1992. His current
research interests include nonlinear phenomena in optical fibers and photonic
bandgap structures, ultrafast optics, and the nonlinear dynamics of coupled
lasers.
Dr. Winful is a fellow of the American Physical Society, and the Optical
Society of America. He received numerous teaching awards, including the
State of Michigan Teaching Excellence Award, the Arthur F. Thurnau Professorship,
the Amoco/University Teaching Excellence Award, the College
of Engineering Teaching Excellence Award, the EECS Professor of the Year
Award (twice), the Tau Beta Pi College of Engineering Outstanding Professor
Award, the EECS Teaching Excellence Award, the EECS Department
Outstanding Achievement Award and the College of Engineering Service
Excellence Award in 2014.
\end{IEEEbiographynophoto}

\begin{IEEEbiographynophoto}{J. Nathan Kutz}
J. Nathan Kutz received the B.S. degree in physics
and mathematics from the University of Washington,
Seattle, WA, USA, in 1990, and the Ph.D. degree in
applied mathematics from Northwestern University,
Evanston, IL, USA, in 1994. He is currently a Professor
of applied mathematics, an Adjunct Professor
of physics and electrical engineering, and a Senior
Data Science Fellow with the eScience Institute,
University of Washington.
\end{IEEEbiographynophoto}





\end{document}